\documentclass[twocolumn]{autart}

\usepackage{amsmath}
\usepackage{mathrsfs}
\usepackage{amsfonts}
\usepackage{graphicx}
\usepackage{amssymb}
\usepackage{latexsym}
\usepackage{array}
\usepackage{ifthen}
\usepackage{wrapfig}
\usepackage{picinpar}
\usepackage{rotating}
\usepackage{morefloats}
\usepackage{enumerate}
\usepackage{bm}
\usepackage{stfloats}
\usepackage{dsfont}
\usepackage[english]{babel}
\usepackage{cite}
\usepackage {enumerate}
\usepackage{subfigure}
\usepackage{color}

\usepackage{algorithm}
\usepackage{algpseudocode}
\usepackage{hyperref}
\usepackage{picins}

\newtheorem{theorem}{Theorem}
\newtheorem{lemma}{Lemma}

\newtheorem{corollary}{Corollary}

\newtheorem{remark}{Remark}

\newtheorem{dproblem}{Design Problem}

\newcommand{\sys}[4]{\left[\begin{array}{c|c}
#1 & #2 \\ \hline #3 & #4
\end{array} \right]}

\def\qed{ \rule{.08in}{.08in}}

\newcommand{\diag}{\mathrm{diag}}
\newcommand{\ess}{\mathrm{ess}}
\newcommand{\ave}{\mathrm{ave}}
\newcommand{\dis}{\mathrm{dis}}

\newcommand{\boldL}{\boldsymbol{L}}
\newcommand{\boldP}{\boldsymbol{P}}
\newcommand{\boldC}{\boldsymbol{C}}
\newcommand{\boldG}{\boldsymbol{G}}

\begin{document}

\begin{frontmatter}

\title{Synchronization of Diverse Agents \\via Phase Analysis\thanksref{footnoteinfo}}
\thanks[footnoteinfo]{This work was supported in part by Hong Hong Research Grants Council under project number GRF 16201120, National Natural Science Foundation of China under grants 62073003 and 72131001, and Shenzhen-Hong Kong-Macau Science and Technology Innovation Fund under number SGDX20201103094600006.}

\author[First]{Dan Wang}
\author[Second]{Wei Chen}
\author[First]{Li Qiu}

\address[First]{Department of Electronic and Computer Engineering, The Hong Kong University of Science and Technology, Clear Water Bay, Kowloon, Hong Kong, China (email: dwangah@connect.ust.hk, eeqiu@ust.hk).}
\address[Second]{Department of Mechanics and Engineering Science \& State Key Laboratory for Turbulence and Complex Systems, Peking University, Beijing 100871, China (email: w.chen@pku.edu.cn).}
%


\begin{abstract}
In this paper, the synchronization of heterogeneous agents interacting over a dynamical network is studied. The edge dynamics can model the inter-agent communications which are often heterogeneous by nature. They can also model the controllers of the agents which may be different for each agent or uniform for all the agents. Novel synchronization conditions are obtained for both cases from a phase perspective by exploiting a recently developed small phase theorem. The conditions scale well with the network and reveal the trade-off between the phases of node dynamics and edge dynamics. We also study the synchronizability problem which aims to characterize the allowable diversity of the agents for which controllers can be designed so as to achieve synchronization. The allowable diversity is captured in terms of phase conditions engaging the residue matrices of the agents at their persistent modes. Controller design algorithms are provided for the cases of agent-dependent and uniform controllers, respectively.
\end{abstract}

\end{frontmatter}
\section{Introduction}

Consensus, agreement, rendezvous, flocking, swarming, and synchronization problems for multi-agent systems have been extensively studied for over two decades. See textbooks and research monographs \cite{RenBeard08,Lin08,BulloCortesMartinez,Mesbahi,Lewis14,Bullo}.
These problems can be unified, more or less, into an output synchronization problem where a number of dynamical agents are connected by a network protocol so that the agents' behaviors are coordinated, e.g., made to converge to a common trajectory.

Generally speaking, there are three main factors involved in synchronization, i.e., the agent dynamics, controllers and inter-agent interactions. Early investigations started with analysis of synchronization, namely, find conditions under which a given set of agents reach synchronization with given controllers and given network. In the simplest case of consensus of a set of identical single integrators via a uniform static controller, a well-known result is that consensus can be reached if and only if the network has a spanning tree \cite{SaberMurray04,RenBeard05TAC}. This discovery has spurred enduring interest in finding graph-theoretic conditions required for consensus under various interaction protocols, such as switching networks, time-varying networks, gossip-type interactions, stochastic communication, etc. See \cite{Mesbahi,Bullo} and the references therein for more details.

On a different direction, researchers have paid attention to synchronization of agents with more complicated dynamics, from linear time-invariant (LTI) systems to time varying nonlinear systems, from homogeneous agents to heterogeneous agents. For those cases, very often graph-theoretic conditions alone are no longer sufficient for ensuring synchronization. Instead, the interplay and trade-off between agent dynamics, controllers and inter-agent interactions become more essential. It is well recognized that the synchronization problem is in essence a feedback stability problem on a disagreement subspace which then invites application of the stability theory.
The small gain theorem is one of the most used results on feedback stability. However, it cannot be directly applied to the synchronization problem due to the persistent modes of the agents which determine the common trajectory that the agents will synchronize to. The theories of positive real systems and negative imaginary systems can overcome this difficulty and have been used to derive synchronization conditions \cite{Allgower2014, HaraPassivity11, NIConsensus}. There have been other insightful efforts in establishing synchronization conditions, for instance \cite{LestasVinnicombe10} by using the notion of S-hull and \cite{Khong16} by using the integral quadratic constraints.

Compared to the analysis, the synthesis problem appears more challenging. There are two major issues here. The first one is to characterize the solvability condition, i.e., whether there exist controllers so that synchronization can be achieved (assuming network connectivity). The second one is to give a design procedure which produces synchronizing controllers. In the simplest case when all the agents are identical integrators, the problem
can be solved by a uniform static network controller \cite{Morse2003,SaberMurray04,RenBeard05TAC,RenBeard05ACC,LinFrancisMaggiore06,SaberMurray}. A more general case is when the agents are general LTI but homogeneous \cite{FaxMurray,SaberMurray,Ma,LiDuanChenHuang,YouXie,GuLewis}. Most of the research efforts adopt the natural strategy of applying a uniform controller to all the agents. Requiring the controllers to be uniform makes the design problem significantly more challenging. The solvability of the problem becomes highly nontrivial \cite{YouXie,GuLewis}. Mathematically these solvability conditions can be removed if the controllers are allowed to be different. However, doing this vastly increases the cost of design and implementation.

In recent years, we see more and more works on the case when the agents are heterogeneous \cite{LestasVinnicombe10, WielandAllgower, Allgower2014,Khong16,LuLiuFeng,WCQ20}. How to characterize the allowable diversity in the agents to ensure the solvability becomes a rather challenging task. One way to handle the differences among the agents is to treat them as model uncertainties or perturbations, described in terms of relative uncertainty, gap metric, integral quadratic constraint, etc., to name a few \cite{Hara_glocal,gu_qiu,Khong16}. Robust control techniques can then be employed to design the controllers so that the model uncertainties or perturbations can be tolerated when the synchronization is achieved. When a uniform controller is adopted, the extent of the uncertainties, i.e., the diversity among the agents allowable to ensure problem solvability according to robust control, is rather small. Other powerful tools are needed when the diversity among the agents is large.

The phase theory recently developed in \cite{CWKQcdc19,CWKQ2021} provides a new perspective in investigating the synchronization problem. It turns out particularly useful in characterizing the large diversity among the agents. With this new perspective, we establish a collection of novel results in this paper that add to the knowledge of synchronization. We draw motivation from real applications, in particular from power networks and unmanned system networks. The agents in these networks, such as generators, loads, and UAVs, often have similar phase properties regardless of their physical sizes. We are also inspired from the early investigation of consensus of integrators with different gains, i.e., integrators multiplied by different scalars \cite{SaberMurray04}. It turns out that as long as the scalar multiplications are all positive, consensus is reached regardless of scalar magnitudes. This hints at a natural way to capture the allowable diversity among the agents ensuring synchronizability in terms of phase bounded cones. This hint is evidenced and further strengthened in later works using positive real and/or negative imaginary theories \cite{Allgower2014, HaraPassivity11, NIConsensus}. While positive realness and negative imaginariness only give qualitative phase descriptions, the newly developed phase theory
opens a wide path for a quantitative study of the trade-off between the phases of agents, controllers, and networks.

At the core of the phase theory is a beautiful formulation of small phase theorem, in which lies the soul of the results developed in this paper. Both analysis and synthesis problems are investigated. For the analysis part, led by the small phase theorem, we obtain scalable synchronization conditions unraveling the trade-off between the phases of node dynamics and edge dynamics. For the synthesis part, guided by the principle of phase compensation,
we characterize the allowable diversity among the agents ensuring the problem solvability in terms of phase bounded cones. Controller design algorithms are provided for the cases of agent-dependent and uniform controllers, respectively.

This paper has made significant new contributions compared to our previous conference paper \cite{WCQ20}, where only the analysis problem with nonuniform communication dynamics has been considered. Theorem \ref{thm:undirected} was stated in \cite{WCQ20} without a proof. The proof is now included. The results on the analysis problem with agent-dependent controllers (Theorem 2) and the synthesis problem (Theorems 3 and 4) are all new and go much beyond the the scope of study in the conference paper.


The rest of this paper is organized as follows. The synchronization problem is formulated in Section~\ref{setup}. Some preliminaries on the phase theory, particularly, the small phase theorem, are given in Section~\ref{phase theory}. Section~\ref{sec: main} and Section \ref{sec: design} present our main results on analysis and synthesis problem, respectively. Section \ref{sec: simulation} illustrates a simulation example. Section~\ref{conclusion} concludes this paper.

{\em Notation and graph basics:} We denote by $\mathbb{R}$ and $\mathbb{C}$ the sets of real and complex numbers. We use $\rho(C)$ and $\overline{\sigma}(C)$ to denote the spectral radius and the largest singular value of the matrix $C$ respectively. The Kronecker product of two matrices $A$ and $B$ is denoted by $A\otimes B$. We use $\mathbf{1}_n$ to denote the $n$-dimensional vector with all its entries equal to $1$. The symbol $\mathrm{diag}\{\cdot\}$ denotes the diagonal operation.


A graph, denoted by $\mathbb{G}\!=\!(\mathcal{V}, \mathcal{E})$, consists of a set of nodes $\mathcal{V}\!=\!\{1,\dots,n\}$ and a set of edges $\mathcal{E}\!=\!\{e_1,\dots,e_l\}$. We also use $(i,j)$ to represent the edge directed from node $i$ to node $j$, where $j$ is called the head and $i$ is called the tail of the edge.
A path from node $i_1$ to node $i_k$ is a sequence of edges $(i_1, i_2), (i_2, i_3),\dots,(i_{k-1}, i_k)$ with $(i_j, i_{j+1})\in \mathcal{E}$ for $j\in \{1,\dots,k-1\}$. A node is called a root if it has paths to all the other nodes in the graph. A graph $\mathbb{G}$ is said to have a spanning tree if it has a root. Furthermore, $\mathbb{G}$ is said to be strongly connected if every node is a root. A graph is undirected if $(i,j)\in \mathcal{E}$ implies $(j,i)\in\mathcal{E}$.

A weighted graph is a graph with each edge associated with a weight. The edge weights are assumed to be positive. Denote by $a_{ji}$ the weight of edge $(i,j)$, where $a_{ji}$ is understood to be zero when there is no edge from node $i$ to $j$. The indegree and outdegree of node $i$ are given by $d_{\mathrm{in}}(i)=\sum_{j=1}^n a_{ij}$ and $d_{\mathrm{out}}(i)=\sum_{j=1}^n a_{ji}$ respectively. A graph is said to be weight-balanced if $d_{\mathrm{in}}(i)=d_{\mathrm{out}}(i)$ for all $i\in\mathcal{V}$. For a weighted graph, its Laplacian matrix $L=[l_{ij}]$ is defined as
\begin{align*}
l_{ij}=\begin{cases}
-a_{ij}, & i\neq j,\\
\sum_{j=1,j\neq i}^n a_{ij}, &i=j.
\end{cases}
\end{align*}
The Laplacian matrix $L$ has all its eigenvalues in the closed right half plane. Also, it has a zero eigenvalue with a corresponding eigenvector being $\mathbf{1}_n$. Furthermore, $0$ is a simple eigenvalue of $L$ if and only if $\mathbb{G}$ has a spanning tree. See \cite{laplacian-survey} for a survey on Laplacian matrices.

\section{Problem Formulation}
\label{setup}
Suppose there are $n$ agents. The dynamics of agent $i$ is given by
\begin{align*}
\dot{x}_i(t)&=A_i x_i(t) + B_iu_i(t), \ \ \ x_i(0)=x_{0i},\\
y_i(t)&=C_i x_i(t),
\end{align*}
where $x_i(t), u_i(t), y_i(t)$ are the state, input, and output of agent $i$, and $A_i,B_i,C_i$ are real matrices. The inputs and outputs of all the agents are of the same dimension $m$ but their states can have different dimensions. Assume that $(A_i,B_i)$ are stabilizable and $(C_i,A_i)$ are detectable for $i=1,\dots,n$.

We are interested in the problem of making the agents converge to a common bounded trajectory generated by a set of persistent modes on the imaginary axis $j\Omega=\{0, \pm j\omega_1, \dots, \pm j\omega_q\}$ through local coordinations among their neighbours. To this end, let $A_i,i=1,\dots,n$ be such that they
share exactly the same set of eigenvalues $j\Omega$ on the imaginary axis and have all the other eigenvalues in the open left half plane. Moreover, each imaginary-axis eigenvalue is semi-simple, i.e., with the same algebraic and geometric multiplicities, and has multiplicity $m$.

Let $P_i(s)=C_i(sI-A_i)^{-1}B_i,i=1,\dots,n$ be the transfer function matrices of the agents. Using partial fractional expansion, one can write $P_i(s)$ into the form
\begin{equation}
\begin{split}
P_i(s)=&\frac{M_{0i}}{s}+\frac{M_{1i}}{s-j\omega_1}+\frac{\bar{M}_{1i}}{s+j\omega_1}+\cdots\\
&+ \frac{M_{qi}}{s-j\omega_q}+\frac{\bar{M}_{qi}}{s+j\omega_q}+\Delta_i(s),\label{agent}
\end{split}
\end{equation}
where $M_{0i}\!=\!\lim\limits_{s\rightarrow 0}sP_i(s)\!\in\! \mathbb{R}^{m\times m}$ and $M_{ki}\!=\!\lim\limits_{s\!\rightarrow j\omega_k}(s\!-\!j\omega_k)P_i(s) \!\in\! \mathbb{C}^{m\times m},k=1,\dots,q$ can be all different for distinct agents, $\bar{M}_{ki}$ denotes the elementwise conjugate of $M_{ki}$, and $\Delta_i(s)$ are all stable but can be diversely different.
Under the earlier assumptions, these residue matrices $M_{ki},k=0,1,\dots,q,i=1,2,\dots,n$, are nonsingular.

The agents exchange information with their neighbours over a graph $\mathbb{G}=(\mathcal{V}, \mathcal{E})$ through the following synchronization protocol:
\begin{align}
u_i(s)&=\sum_{(j,i)\in\mathcal{E}}a_{ij}C_{ij}(s)(y_j(s)-y_i(s)),\quad i\in\mathcal{V},\label{protocol}
\end{align}
where $a_{ij}>0$ are feedback gains ($a_{ij}$ is understood to be $0$ if $(j,i)\!\notin\! \mathcal{E}$) and $C_{ij}(s)$, assumed to be $m\times m$ stable transfer matrices, represent the feedback dynamics. The feedback dynamics $C_{ij}(s)$ can model the communication imperfections in each interconnection which are likely all different. They can also model the controllers applied to the agents which may be different for each agent, or uniform through all of the agents. We assume throughout the paper that $\mathbb{G}$ has a spanning tree.

The agents are said to achieve synchronization if
\begin{align*}
|y_i(t)-y_j(t)|\rightarrow 0 \text{ as }t\rightarrow\infty
\end{align*}
for all $i,j\in \mathcal{V}, i\neq j$ for any initial conditions. Further, if the agents have only one common unstable pole at the origin, their outputs will converge to a common constant trajectory; if the agents have common pairs of complex poles on the imaginary axis, their outputs will converge to a common oscillating trajectory. The former case is well known as the consensus problem.


Denote by
\begin{align*}
\boldsymbol{P}(s) &= \mathrm{diag}\{P_1(s),\dots, P_n(s)\}, \ \ \ \ x_0=\begin{bmatrix}x_{01}'& \cdots & x_{0n}'\end{bmatrix}',\\
u(t)&=\begin{bmatrix}u_1(t)'& \cdots &u_n(t)'\end{bmatrix}',\
y(t)=\begin{bmatrix}y_1(t)'& \cdots & y_n(t)'\end{bmatrix}'.
\end{align*}
From the agents' dynamics, we have in frequency domain
\begin{align}
    y(s)=\boldsymbol{P}(s)u(s)+\boldsymbol{H}(s)x_0,\label{response}
\end{align}
where
\begin{align*}
\boldsymbol{H}(s)=\mathrm{diag}\{C_1,\dots,C_n\}\left(sI-\diag\{A_1,\dots,A_n\}\right)^{-1}.
\end{align*}
On the other hand, the synchronization protocol (\ref{protocol}) can be written into the compact form
\begin{align}
    u(s)=-\boldL(s)y(s),\label{procompact}
\end{align}
where $\boldL(s)=[L_{ij}(s)]$ is a dynamical matrix Laplacian whose $ij$th block is given by
\begin{align*}
L_{ij}(s)=\begin{cases}
-a_{ij}C_{ij}(s), & i\neq j,\\
\sum_{j=1,j\neq i}^n a_{ij}C_{ij}(s), &i=j.
\end{cases}
\end{align*}
Note that $\boldL(j\omega)$ has $m$ zero eigenvalues with corresponding eigenvectors being $\mathbf{1}_n\otimes x$ for all $\omega\in\mathbb{R}\cup \{\infty\}$, where $x\!\in\!\mathbb{C}^m$ is an arbitrary nonzero vector.

The closed-loop synchronization dynamics is now
represented by the block diagram in Figure \ref{network_framework}.
Moreover, we have (by substituting (\ref{procompact}) into (\ref{response}))
\begin{align}
    y(s)=\left(I+\boldsymbol{P}(s)\boldL(s)\right)^{-1}\boldsymbol{H}(s)x_0.\label{closed}
\end{align}

\setlength{\unitlength}{1.25mm}
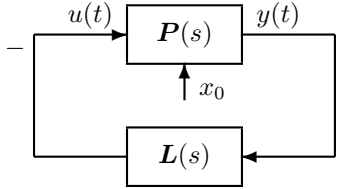
\begin{figure}[htb]
\begin{center}
\begin{picture}(36,20)
\thicklines
\put(3,16){\vector(1,0){10}}
\put(13,13){\framebox(12,6){$\boldsymbol{P}(s)$}}
\put(19,9){\vector(0,1){4}}
\put(20,8){\makebox(4,4){$x_0$}}
\put(25,16){\line(1,0){10}}
\put(35,16){\line(0,-1){13}}
\put(35,3){\vector(-1,0){10}}
\put(13,0){\framebox(12,6){$\boldL(s)$}}
\put(13,3){\line(-1,0){10}}
\put(3,3){\line(0,1){13}}
\put(7,16){\makebox(4,4){$\displaystyle u(t)$}}
\put(27,16){\makebox(4,4){$\displaystyle y(t)$}}
\put(0,13){\makebox(2,3){$-$}}
\end{picture}
\caption{Block diagram of synchronization problem.}
\label{network_framework}
\end{center}
\end{figure}

Such a synchronization problem can be transformed to a feedback stability problem on a disagreement subspace. We define the average output trajectory to be
\[
y^{\ave}(t)=\frac{1}{n}(\mathbf{1}_n\otimes I_m)' y(t)
\]
and the output disagreement to be
\[
y^{\dis}(t)=y(t)-(\mathbf{1}_n\otimes y^{\ave}(t)).
\]
Synchronization is achieved if and only if $y^{\dis}(t)\rightarrow 0$ as $t\rightarrow \infty$ and $y^{\ave}(t)$ converges to a bounded trajectory described by the persistent modes in $j\Omega$.

Let $Q$ be such that $\begin{bmatrix}\frac{\mathbf{1}_n}{\sqrt{n}}&Q\end{bmatrix}$ is an orthogonal matrix. We have
\begin{align*}
   &I+\boldsymbol{P}(s)\boldL(s)=\left(\!\begin{bmatrix}\frac{\mathbf{1}_n}{\sqrt{n}}&Q\end{bmatrix}\!\otimes\! I_m\!\right)\\
   &\cdot \begin{bmatrix}
   I& (\frac{\mathbf{1}'_n}{\sqrt{n}}\otimes I_m)\boldsymbol{P}(s)\boldL(s)(Q\otimes I_m)\\ 0 & I\!+\!(Q'\!\otimes\! I_m)\boldsymbol{P}(s)\boldL(s) (Q\!\otimes\! I_m) \end{bmatrix}\!\!
    \left(\!\begin{bmatrix}\frac{\mathbf{1}_n}{\sqrt{n}}&Q\end{bmatrix}\!\otimes\! I_m\!\right)'.
\end{align*}
Then $(I\!+\!\boldsymbol{P}(s)\boldL(s))^{-1}$ is given by
\begin{align*}
&(I+\boldsymbol{P}(s)\boldL(s))^{-1}=\left(\begin{bmatrix}\frac{\mathbf{1}_n}{\sqrt{n}}&Q\end{bmatrix}\otimes I_m\right)\\
&\cdot \!\!\begin{bmatrix}
I& \!-\!\left(\!\frac{\mathbf{1}'_n}{\sqrt{n}}\!\otimes\! I_m\!\right)\!\!\boldsymbol{P}(s)\boldL(s)(Q\!\otimes\! I_m)S(s)\\ 0 & S(s) \end{bmatrix}\!\!\!
\left(\!\begin{bmatrix}\frac{\mathbf{1}_n}{\sqrt{n}}&Q\end{bmatrix}'\!\!\otimes\! I_m\!\right)\!,
\end{align*}
where
\begin{equation*}
S(s)= {\Big(}I+(Q'\otimes I_m)\boldsymbol{P}(s)\boldL(s)(Q\otimes I_m){\Big)}^{-1}.
\end{equation*}
In view of (\ref{closed}), one can compute $y^{\ave}(s)$ and $y^{\dis}(s)$ to be
\begin{align*}
y^{\ave}(s)\!=& \left(\mathbf{1}'_n \!\otimes\! I_m\right)\\
&\cdot\left( I\!-\! \boldsymbol{P}(s)\boldL(s)(Q\!\otimes\! I_m)S(s)(Q'\!\otimes\! I_m)\right)\!
\boldsymbol{H}(s)x_0,\\
y^{\dis}(s)\!=&(Q\otimes I_m)S(s)(Q'\otimes I_m)\boldsymbol{H}(s)x_0.
\end{align*}

Let us have a closer look at $y^{\dis}(s)$. The sensitivity function $S(s)$ is said to be internally stable if it belongs to $\mathcal{RH}_{\infty}$ and the unstable poles $j\Omega$ of $(Q'\otimes I_m)\boldP(s)$, each with multiplicity $(n-1)m$, remain as the poles of $(Q'\otimes I_m)\boldP(s)\boldL(s)(Q\otimes I_m)$ with the same multiplicity. The latter requires that the rank of
\[
(Q'\otimes I_m)\diag\{M_{k1},\dots, M_{kn}\}\boldL(j\omega_k)(Q\otimes I_m)
\]
is equal to $(n-1)m$ for all $k=0,\dots, q$. If $S(s)$ is internally stable, then $S(j\omega_k)=0$ and $y^{\dis}(s)$ has no pole at $j\Omega$. Hence, $y^{\dis}(t)$ converge to zero as $t\rightarrow \infty$.


Now we analyze $y^{\ave}(s)$. Under the assumption that $S(s)$ is internally stable, we have $S(j\omega_k)=0$, which implies
\[
I\!-\! \boldsymbol{P}(s)\boldL(s)(Q\!\otimes\! I_m)S(s)(Q'\!\otimes\! I_m)\in \mathcal{RH}_{\infty}.
\]
Therefore, the steady state of $y^{\ave}(t)$ contains exactly those modes in $j\Omega$.

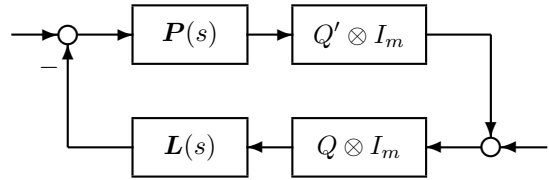
\begin{figure}[htb]
\begin{center}
\begin{picture}(58,20)
\thicklines
\put(0,15){\vector(1,0){5}}
\put(6,15){\circle{2}}
\put(7,15){\vector(1,0){6}}
\put(13,12){\framebox(12,6){$\displaystyle \boldsymbol{P}(s)$}}
\put(25,15){\vector(1,0){5}}
\put(30,12){\framebox(14,6){$\displaystyle Q'\otimes I_m$}}
\put(44,15){\line(1,0){7}}
\put(51,15){\vector(0,-1){11}}
\put(51,3){\circle{2}}
\put(57,3){\vector(-1,0){5}}
\put(50,3){\vector(-1,0){6}}
\put(30,0){\framebox(14,6){$\displaystyle Q\otimes I_m$}}
\put(30,3){\vector(-1,0){5}}
\put(13,0){\framebox(12,6){$\displaystyle \boldL(s)$}}
\put(13,3){\line(-1,0){7}}
\put(6,3){\vector(0,1){11}}
\put(3,10){\makebox(2,3){$-$}}
\end{picture}
\caption{Feedback stability problem.}
\label{synchronization-stability}
\end{center}
\end{figure}

Combining the above analysis, to reach synchronization, $S(s)$ should be internally stable, which is equivalent to the internal stability of the closed-loop system shown in Figure \ref{synchronization-stability}.


Now that synchronization is connected to feedback stability so that results from stability theory can be applied.
The purpose of this paper is to add to the understandings of synchronization by exploiting the recently developed phase theory for multi-input multi-output (MIMO) LTI systems. A brief review will be provided later. Particularly, the small phase theorem formulated in \cite{CWKQ2021} enacts a quantitative phasic approach in studying the synchronization problems. We will see that the phase notion offers great advantages in incorporating the diverse heterogeneity in the networks, leading to a collection of novel results and new understandings.

We study both the analysis and synthesis problems. The analysis problem aims to seek conditions for a given set of heterogeneous agents achieving synchronization under given network topology and given edge dynamics. Two scenarios are considered: 1) the edge dynamics model the communication imperfections among the agents; 2) the edge dynamics model the controllers of the agents. The synthesis problem aims at designing controllers, if possible, so that a given set of heterogeneous agents reach synchronization under given network topology. Two main issues need to be addressed. The first one is to characterize the solvability
condition with respect to the diversity of the agents. The second one is to give controller design algorithms when the problem is solvable. We consider the cases of agent-dependent controllers and a uniform controller, respectively.

\section{Phases of Matrices and LTI MIMO Systems}
\label{phase theory}
\subsection{Phases and essential phases of matrices}
\label{sec:matrix phase}
In this section, we review some basics of phases of semi-sectorial matrices and essential phases of essentially semi-sectorial matrices \cite{WCKQ20,CWKQ2021,QWMC22}.
We will see later that these notions play an important role in the study of stability and synchronization of dynamical networks.

The numerical range of a matrix $C \in \mathbb{C}^{n\times n}$ is defined to be $W(C) = \{ x^*Cx: x \in \mathbb{C}^n, \|x\|=1\}$. It is a compact and convex subset of
$\mathbb{C}$, and contains the spectrum of $C$ \cite{horntopics}.
A matrix $C$ is said to be sectorial if $0\notin W(C)$, or equivalently, $W(C)$ is contained in an open half complex plane. A sectorial $C$ has a sectorial decomposition
\begin{align}
C=T^*DT \label{sd}
\end{align}
for some nonsingular $T$ and diagonal unitary $D$, where $D$ is unique up to a permutation and has its eigenvalues distributed in an arc on the unit circle with length less than $\pi$ \cite{ZhangFuzhen2015}.
The phases of $C$, denoted by
\[
\overline{\phi}(C)=\phi_1(C)\geq\phi_2(C)\geq\dots\geq\phi_n(C)=\underline{\phi}(C),
\]
are defined as the phases of the eigenvalues of $D$ so that $\overline{\phi}(C)-\underline{\phi}(C) \!<\!\pi$. The phases are multi-valued functions. Their values are only uniquely determined after we make a selection of $\displaystyle \gamma(C) = [\overline{\phi}(C)+\underline{\phi}(C)]/2$, called the phase center of $C$, in $\mathbb{R}$. The phases are said to take the principal values if $\gamma (C)$ is selected in $(-\pi, \pi]$. However, we will not always select $\gamma(C)$ using its principal value. Following the standard way of selecting the phase of a complex scalar, we will select $\gamma(C)$ to make it continuous in the elements of $C$. In this way the phases are continuous in the elements of $C$. The phases defined in this way resemble the canonical angles of $C$ introduced in \cite{FurtadoJohnson2001}. Denote
\[\phi (C) = \begin{bmatrix} \phi_1 (C) & \phi_2 (C) & \cdots & \phi_n(C) \end{bmatrix}\]
and
\[
\Phi(C)=[\underline{\phi}(C),\overline{\phi}(C)].
\]

\begin{figure}[htb]
    \centering
    \includegraphics[scale=0.45]{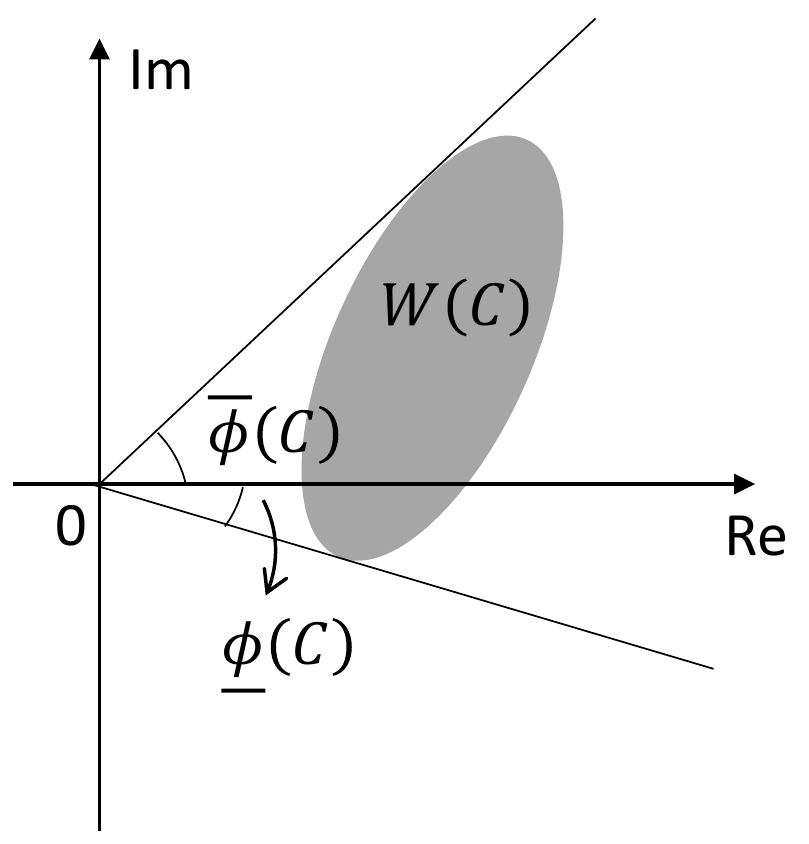}
    \vspace{-5pt}
    \caption{Geometric interpretations of $\overline{\phi}(C)$ and $\underline{\phi}(C)$.}
    \label{fig:sectorial}
\end{figure}

A graphic illustration of phases is depicted in Figure \ref{fig:sectorial}. The two angles from the positive real axis to the two supporting rays of $W(C)$ are $\overline\phi(C)$ and $\underline\phi(C)$ respectively. The other phases of $C$ lie in between.

The phase definition can be extended to a broader class of matrices. A matrix $C$ is said to be semi-sectorial if $0$ is not in the interior of $W(C)$, or equivalently, $W(C)$ is contained in a closed half plane. A degenerate case of such matrices is when the numerical range has no interior and is given by a straight interval containing the origin in its relative interior. In this case, $C$ is in fact a rotated Hermitian matrix and has the decomposition
\[
C= T^* \mathrm{diag} \{ 0_{n-r}, e^{j(\gamma(C)+\pi/2)} I,  e^{j(\gamma(C)-\pi/2)} I \} T,
\]
where $r\!=\!\mathrm{rank}(C)$. The value of $\gamma(C)$ is determined modulo $\pi$ and there are two possible principal values in $(-\pi, \pi]$. The phases of $C$ are several copies of $\gamma(C)+\pi/2$
and several copies of $\gamma(C)-\pi/2$.

\begin{figure}[htb]
\centering
\subfigure[A semi-sectorial matrix]{
\begin{minipage}{0.23\textwidth}\centering
\includegraphics[scale=0.4]{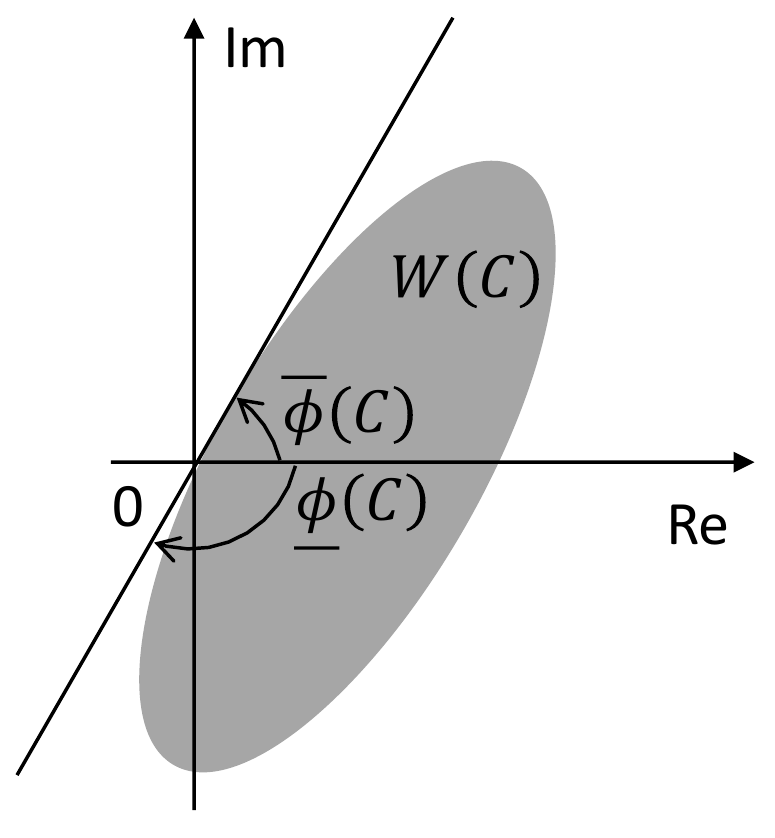}
\end{minipage}}
\subfigure[A quasi-sectorial matrix]{
\begin{minipage}{0.23\textwidth}\centering
\includegraphics[scale=0.4]{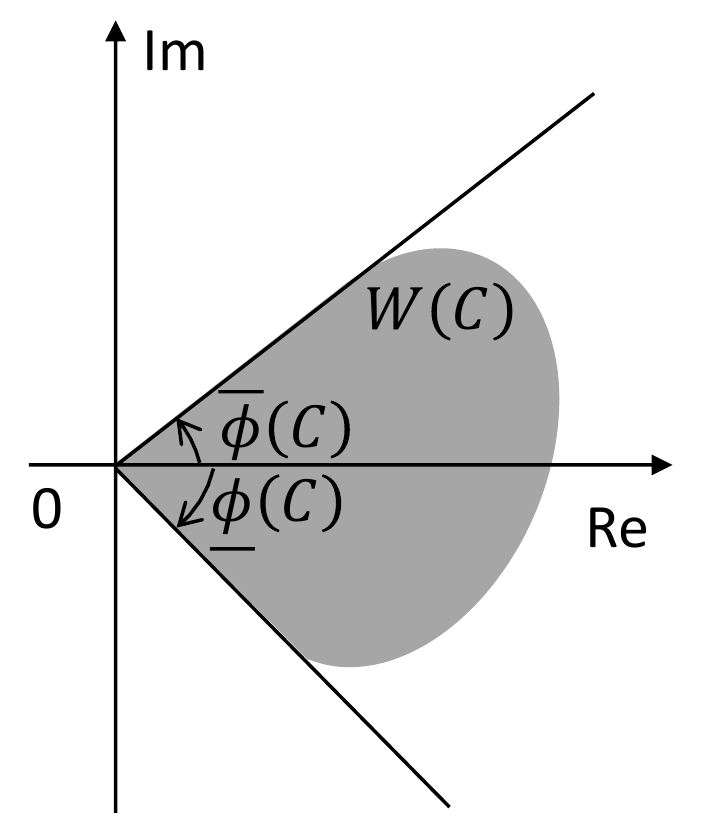}
\end{minipage}}
\caption{Semi-sectorial and quasi-sectorial matrices.}
\label{fig:semisec}
\end{figure}

The generic case of semi-sectorial matrices is when the numerical range has nonempty interior. If $C$ is nonsingular, then $C$ admits the following decomposition \cite{FurtadoJohnson2003}
\begin{equation}\label{gsd}
C=T^*\begin{bmatrix} D & 0 \\  0 & E \end{bmatrix}T,
\end{equation}
where $T$ is nonsingular, $D=\mathrm{diag} \{ e^{j\theta_1}, \dots, e^{j\theta_m}\}$
with $\theta_0+\pi/2 \geq \theta_1  \geq \cdots \geq \theta_m \geq \theta_0-\pi/2$, and
\[
E=\mathrm{diag} \left\{ e^{j\theta_0} \begin{bmatrix} 1 & 2 \\ 0 & 1 \end{bmatrix}, \dots , e^{j\theta_0} \begin{bmatrix} 1 & 2 \\ 0 & 1 \end{bmatrix} \right\}.
\]
The phases of $C$ are defined to be $\theta_1, \dots, \theta_m$ along with $(n-m)/2$ copies of $\theta_0\pm\pi/2$ so that $\overline{\phi}(C)-\underline{\phi}(C) \!\leq\! \pi$. If $C$ is singular, then $0$ is an eigenvalue on the boundary of $W(C)$ which has to be a normal eigenvalue \cite{horntopics}. From the definition of normal eigenvalue, it follows that $C$ has the following decomposition
\begin{align}
C=U\begin{bmatrix}0&0\\0&\hat{C}\end{bmatrix}U^*,  \label{semidec}
\end{align}
where $U$ is unitary and $\hat{C}\in\mathbb{C}^{r\times r}$ is full rank. If further, $0$ is on the boundary of $W(\hat{C})$, then $\hat{C}$ has a decomposition in the form of (\ref{gsd}); otherwise, $0\!\notin\! W(\hat{C})$ and consequently $\hat{C}$ has a decomposition in the form of (\ref{sd}). In either case, the phases of $C$ are defined to be the phases of $\hat{C}$. Note that in the former case, $0$ is on the smooth boundary of $W(C)$ while in the latter case, $0$ is a sharp point of $W(C)$. The latter case has its own interest as it defines a subset of semi-sectorial matrices that are congruent to the direct sum of $0$ and sectorial matrices. We call such matrices quasi-sectorial matrices. A graphic illustration of semi-sectorial and quasi-sectorial matrices is depicted in Figure \ref{fig:semisec}.


The matrix phases have many nice properties; see \cite{WCKQ20,CWKQ2021}. Here we briefly review several of them which will be useful in later developments.

The first property is concerned with the phases of matrix compressions. Let $C\in\mathbb{C}^{n\times n}$ be a semi-sectorial matrix. Then $\tilde{C}\!=\!X^*CX$, where $X\!\in\!\mathbb{C}^{n\times k}$ has full column rank, is said to be a compression of $C$.
\begin{lemma}[\!\!\cite{FurtadoJohnson2003}]
\label{lemma:semi-compression}
Let $C\!\!\in\!\!\mathbb{C}^{n\times n}$ be nonzero semi-sectorial and $\tilde{C}$ be a nonzero compression of $C$. Then $\tilde{C}$ is semi-sectorial and
$
\underline\phi(C) \leq \underline\phi(\tilde{C}) \leq \overline\phi(\tilde{C}) \leq \overline\phi(C).
$
\end{lemma}

The next lemma discusses the product of a semi-sectorial matrix and a quasi-sectorial matrix.
\begin{lemma}[\!\!\cite{QWMC22}]
\label{lemma:semi-major}
Let $A, B\in\mathbb{C}^{n\times n}$ be quasi-sectorial and semi-sectorial with phase centers $\gamma(A)$ and $\gamma(B)$ respectively. Then $AB$ has $\mathrm{rank}(AB)^2$ nonzero eigenvalues $\lambda_i(AB), i=1,\dots,\mathrm{rank}(AB)^2$, and $\angle \lambda_i(AB)$ can take value in $(\gamma(A)+\gamma(B)-\pi, \gamma(A)+\gamma(B)+\pi)$. Moreover,
\begin{align*}
\underline\phi(A)+\underline\phi(B) \leq \angle\lambda_i(AB) \leq \overline\phi(A)+\overline\phi(B).
\end{align*}
\end{lemma}

Another useful property concerns the phases of the Kronecker product of two semi-sectorial matrices. See the next lemma, a simple extension of Theorem 11.1 in \cite{WCKQ20}.
\begin{lemma}
\label{thm: kronecker}
Let $A\!\in\!\mathbb{C}^{n\times n},B\!\in\!\mathbb{C}^{m\times m}$ be semi-sectorial. If $\overline\phi(A) + \overline\phi(B) - \underline\phi(A) - \underline\phi(B) \leq \pi$, then $A\otimes B$ is semi-sectorial and has $\mathrm{rank}(A)\mathrm{rank}(B)$ number of phases given by
$\phi_i(A) \!+\! \phi_j(B), 1 \!\leq\! i \!\leq\! \mathrm{rank}(A), 1 \!\leq\! j \!\leq\! \mathrm{rank}(B)$.
\end{lemma}

In many applications, we may encounter a matrix which is not necessarily semi-sectorial but can be made semi-sectorial by diagonal similarity transformation. Such a matrix is said to be essentially semi-sectorial. For such a matrix $C$, we define its (largest and smallest) essential phases to be
\begin{align*}
\overline\phi_{\mathrm{ess}}(C)\!=\!\inf_{D\in\mathcal{D}} \overline\phi(D^{-1}CD),\ \
\underline\phi_{\mathrm{ess}}(C)\!=\!\sup_{D\in\mathcal{D}} \underline \phi(D^{-1}CD),
\end{align*}
where $\mathcal{D}$ is the set of positive definite diagonal matrices. Here the infimum and supremum are taken over $D\in\mathcal{D}$
such that $D^{-1}CD$ is semi-sectorial and $\overline\phi(D^{-1}CD)$ and $\underline \phi(D^{-1}CD)$ take their principal values. Such an essential phase problem is reminiscent of the essential gain problem that one may find more familiar with. The essential gain of a matrix $C$ is defined as
\[
\overline\sigma_{\mathrm{ess}}(C)=\inf_{D\in\mathcal{D}} \overline\sigma(D^{-1}CD),
\]
which has proven useful in various applications. It has been studied in \cite{Safonov1982} with input from \cite{Bauer}. Moreover, it is a special problem in the so-called ``$\mu$-analysis'' \cite{Doyle1982,Zhou} in robust control.

It is known that the essential gain problem can be solved numerically but does not have an analytic solution in general. In the case of a nonnegative matrix $C$, the essential gain has an analytic expression $\overline\sigma_{\mathrm{ess}}(C)=\rho(C)$ and the optimal scaling matrix $D$ can be obtained from the Perron eigenvectors of $C$ \cite{StoerWitzgall}. It is a similar situation for the essential phase problem. In general the problem can be solved numerically. For some special classes of matrices, analytic or semi-analytic solutions can be obtained.
One may refer to \cite{QWMC22} for more details. Here we briefly review the essential phases of Laplacian matrices which will be useful in later developments.


We first consider the case of strongly connected graphs. In general, the Laplacian matrix $L$ is not semi-sectorial.
Let $v$ be a positive left eigenvector of $L$ corresponding to the zero eigenvalue, i.e., $v'L\!=\!0$. Let $V\!\!=\!\mathrm{diag}\{v\}$ and $D_o\!=\!V^{\!-\!1/2}$.
We have the following result.



\begin{lemma}[\!\cite{QWMC22}]\label{essphaseLap}
Let $\mathbb{G}$ be a strongly connected directed graph and $L$ be its Laplacian matrix. The following statements are equivalent:
\vspace{-5pt}
\begin{enumerate}[(1)]
\item $L$ is quasi-sectorial.
\item $L$ is semi-sectorial.
\item $\mathbb{G}$ is weight-balanced.
\end{enumerate}
Moreover, it holds that
\begin{align*}
\overline\phi_{\mathrm{ess}}(L)&=\overline\phi(D_o^{-1}LD_o)=\overline\phi(VL),\\
\underline\phi_{\mathrm{ess}}(L)&=\underline\phi(D_o^{-1}LD_o)=\underline\phi(VL).
\end{align*}
\end{lemma}

\vspace{-10pt}

Note that $VL$ is a Laplacian matrix with $\mathbf{1}$ being a common left and right eigenvector corresponding to eigenvalue $0$. This means that $VL$ is the Laplacian matrix of a weight-balanced graph.
By Lemma \ref{essphaseLap}, $VL$ is quasi-sectorial. Hence $D_o^{-1}LD_o$ is quasi-sectorial and  $\overline{\phi}_{\mathrm{ess}}(L)<\frac{\pi}{2}$.
It then follows from \cite[Lemma 2.3]{WCKQ20} that
\[
\max_i \{\angle \lambda_i(L)\} \!\leq\!  \overline\phi_{\mathrm{ess}}(L)\!<\!\pi/2.
\]
Since $L$ is real, there holds $\underline \phi_{\mathrm{ess}}(L)\!=\!-\overline\phi_{\mathrm{ess}}(L)$. For
this reason, hereinafter we use $\phi_{\mathrm{ess}}(L)$ to represent $\overline\phi_{\mathrm{ess}}(L)$ for notational simplicity.
In the case of an undirected graph, $L$ is symmetric and hence $\phi_{\mathrm{ess}}(L) = 0$. This suggests the use of $\phi_{\mathrm{ess}}(L)$ as a measure of ``directedness'' of a graph.

We proceed to consider the case where the graph is not strongly connected but has a spanning tree. In this case, one can decompose the graph into multiple strongly connected components. Suppose the graph has $n_1$ roots and $\nu$ strongly connected components. Without loss of generality, one can relabel the nodes to form $\nu$ groups
\begin{align}
\!\{1,\dots,n_1\},\{n_1\!+\!1,\dots,n_2\},\dots,\{n_{\nu-1}\!+\!1,\dots,n\}\label{strongcd}
\end{align}
so that the nodes in each group correspond to a strongly connected component and the first component contains all the roots. The Laplacian $L$ can be written accordingly in the Frobenius normal form \cite{BrualdiRyser}
\begin{equation}
\begin{split}
L=\begin{bmatrix}
L_{11} & 0& \cdots &0\\
L_{21} & L_{22} & \cdots &0\\
\vdots &\vdots &\ddots &\vdots\\
L_{\nu1} & L_{\nu2} &\cdots & L_{\nu\nu} \label{eq: L_decomp}
\end{bmatrix},
\end{split}
\end{equation}
where $L_{11}$ is the Laplacian of the subgraph induced by all the roots and $L_{\kappa\kappa},\kappa=2,\dots,\nu$ are nonsingular M-matrices.
Moreover, $L$ has a nonnegative left eigenvector $v=\begin{bmatrix}v_1'&0\end{bmatrix}'$ corresponding to eigenvalue $0$, where $v_1$ is a positive left eigenvector of $L_{11}$ corresponding to eigenvalue $0$. Since $v$ is not positive, Lemma \ref{essphaseLap} fails to hold in this case.

Nevertheless, as will be seen later, one often needs to find the essential phase of each $L_{\kappa\kappa}$ on the diagonal. Clearly, $\phi_{\mathrm{ess}}(L_{11})$ can be determined as in Lemma \ref{essphaseLap} for $L_{11}$ is the Laplacian associated to the first strongly connected component.
The following lemma shows that $\phi_{\ess}(L_{\kappa\kappa}),\kappa\!=\!2,\dots,\nu$ exist and are bounded by $\phi_{\mathrm{ess}}(\tilde{L}_\kappa),\kappa\!=\!2,\dots,\nu$ respectively, where $\phi_{\ess}(L_{\kappa\kappa})$ represents $\overline{\phi}_{\ess}(L_{\kappa\kappa})$ and $\tilde{L}_\kappa$ is the Laplacian matrix of the $\kappa$th strongly connected component of the graph.

\begin{lemma}[\!\cite{QWMC22}]
\label{epLii}
Let $\mathbb{G}$ be a directed graph with a spanning tree and $L$ be its Laplacian matrix in the form of (\ref{eq: L_decomp}). Then $\phi_{\mathrm{ess}}(L_{\kappa\kappa})\leq \phi_{\mathrm{ess}}(\tilde{L}_\kappa),\kappa=2,\dots,\nu$.
\end{lemma}

\subsection{Phases of MIMO systems and small phase theorem}
\begin{figure}[htb]
\centering
\includegraphics[scale=0.6]{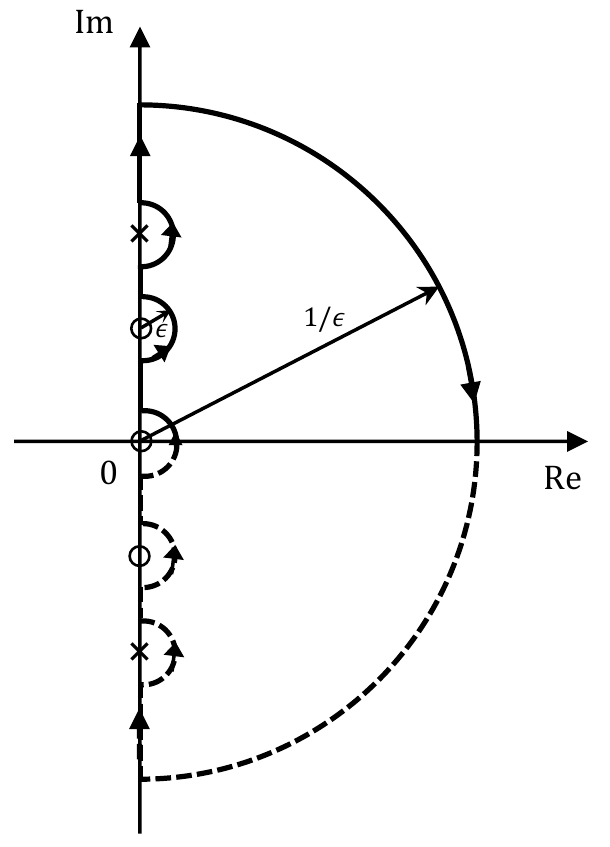}
\vspace{-5pt}
\caption{Indented $j\omega$-axis: ``x'' and ``o'' denote the $j\omega$-axis poles and zeros.}
\label{fig:indent axis}
\end{figure}

Here we review the phase notion of MIMO systems and the small phase theorem recently developed in \cite{CWKQ2021}. Consider an $m\times m$ real rational proper semi-stable system $G$ with $j\Omega$ being the set of poles on the imaginary-axis. Then, $G$ is said to be frequency-wise semi-sectorial if
\begin{enumerate}
    \item $G(j\omega)$ is semi-sectorial for all $\omega \in [-\infty,\infty]\backslash \Omega$;
    \item there exists an $\epsilon^* > 0$ such that for all $\epsilon\leq \epsilon^*$, $G(s)$ has a constant rank and is semi-sectorial along the indented imaginary axis shown in Figure \ref{fig:indent axis}, where the half-circle detours with radius $\epsilon$ are taken at both the poles and finite zeros of $G(s)$ at the frequency axis and a half-circle detour with radius $1/\epsilon$ is taken if infinity is a zero of $G(s)$.
\end{enumerate}

A typical example of the frequency-wise semi-sectorial system is $1/s^n$. For the agents considered in this paper, each pole in $j\Omega$ is at most a simple pole of each element of the transfer function matrix. For such a system $G$, it is frequency-wise semi-sectorial if $G(j\omega)$ is semi-sectorial for all $\omega\in[-\infty, \infty]\backslash \Omega$ and the residue matrix at each pole in $j\Omega$ is semi-sectorial. A stable system $G$ is said to be frequency-wise sectorial if $G(j\omega)$ is sectorial for all $\omega\in[-\infty,\infty]$. Clearly, a frequency-wise sectorial system does not have transmission zeros on the imaginary axis.

For a semi-stable frequency-wise semi-sectorial $G$, its DC phases are defined as $\phi(G(0))$ (if $0$ is neither a pole nor a zero of $G$) or $\phi(G(\epsilon))$ (if $0$ is a pole or a zero of $G$). For simplicity, we assume throughout that $\gamma(G(0))\!=\!0$ (or $\gamma(G(\epsilon))\!=\!0$). Let $\phi(G(s))$, i.e., the vector of phases of $G(s)$, be defined so that
$\gamma(G(s))$ is continuous along the indented imaginary axis. Then $\phi(G(s))$ is an $\mathbb{R}^m$-valued continuous function along the indented imaginary axis, which we call the phase response of $G$. 
For semi-stable frequency-wise semi-sectorial $G$, its maximum and minimum phases are defined to be
\begin{align*}
\overline{\phi}(G)\!=\!\sup_{\omega\in[0,\infty]\backslash\Omega}\overline{\phi}(G(j\omega)),\ \
\underline{\phi}(G)\!=\!\inf_{\omega\in[0,\infty]\backslash\Omega}\underline{\phi}(G(j\omega)).
\end{align*}

The phase notion generalizes the well-known positive realness. In the language of phase, positive real systems are semi-stable frequency-wise semi-sectorial with $[ \underline{\phi}(G), \overline{\phi}(G)] \!\subset\![-\frac{\pi}{2},\frac{\pi}{2}]$, while strongly positive real systems are stable frequency-wise sectorial with $[ \underline{\phi}(G), \overline{\phi}(G)]\subset(-\frac{\pi}{2},\frac{\pi}{2})$.

Consider two $m\times m$ real rational proper transfer function matrices $G$ and $H$. The feedback interconnection of $G$ and $H$, as shown in Figure \ref{fdbk}, is said to
be stable if the Gang of Four matrix
\begin{align*}
G\#H=\begin{bmatrix}
  (I + HG)^{-1} & (I + HG)^{-1}H\\
  G(I + HG)^{-1} & G(I + HG)^{-1}H
\end{bmatrix}
\end{align*}
is proper and stable, i.e., $G\#H \in \mathcal{RH}^{2m \times 2m}_\infty$.

\setlength{\unitlength}{1.23mm}
\begin{figure}[htb]
\begin{center}
\begin{picture}(37,19)
\thicklines
\put(0,15){\vector(1,0){6}}
\put(7,15){\circle{2}}
\put(8,15){\vector(1,0){6}}
\put(14,12){\framebox(8,6){$G$}}
\put(22,15){\line(1,0){7}}
\put(29,15){\vector(0,-1){10}}
\put(28,4){\vector(-1,0){6}}
\put(29,4){\circle{2}}
\put(36,4){\vector(-1,0){6}}
\put(14,1){\framebox(8,6){$H$}}
\put(14,4){\line(-1,0){7}}
\put(7,4){\vector(0,1){10}}
\put(3,8){\makebox(5,5){$y_1$}}
\put(29,7){\makebox(5,5){$y_2$}}
\put(0,14){\makebox(5,5){$w_1$}}
\put(31,-1){\makebox(5,5){$w_2$}}
\put(8,14){\makebox(5,5){$u_1$}}
\put(24,-1){\makebox(5,5){$u_2$}}
\put(6,8){\makebox(6,10){$-$}}
\end{picture}
\vspace{-3pt}
\caption{A standard feedback system.}
\label{fdbk}
\end{center}
\end{figure}
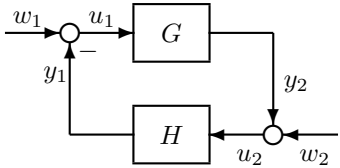


The following small phase theorem, a counterpart of the small gain theorem, was formulated in \cite{CWKQ2021}.
\begin{lemma}[small phase theorem]
\label{thm:gspt}
Let $G$ be semi-stable frequency-wise semi-sectorial with $j\Omega$ being the set of poles on the imaginary axis, where $j\omega_0\in j\Omega$ is at most a simple pole of each element of $G$. Let $H\in\mathcal{RH}_\infty$ be frequency-wise sectorial. Then $G\#H$ is stable if
\begin{align*}
\overline{\phi}(G(j\omega))+\overline\phi(H(j\omega))&<\pi,\\
\underline{\phi}(G(j\omega))+\underline\phi(H(j\omega))&>-\pi 
\end{align*}
for all $\omega\in[0,\infty] \backslash \Omega$.
\end{lemma}

The small phase theorem subsumes a well-known version of passivity theorem \cite{Kottenstette,LiuYao16} which states that $G\#H$ is stable if $G$ is
positive real and $H$ is strongly positive real.

\section{Analysis for Synchronization}\label{sec: main}
\subsection{Synchronization with nonuniform edge dynamics}

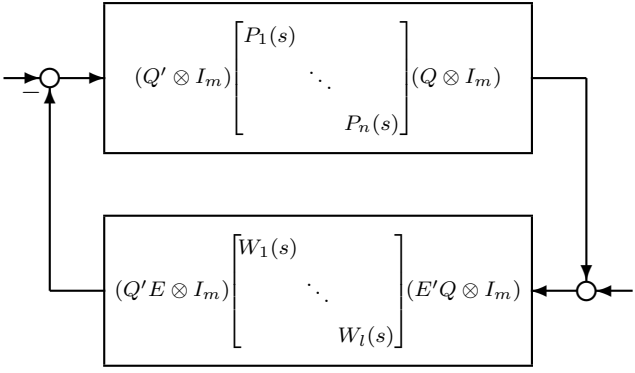
\begin{figure}[htb]
\begin{center}
\begin{picture}(68,40)
\thicklines
\put(0,31){\vector(1,0){4}}
\put(5,31){\circle{2}}
\put(6,31){\vector(1,0){5}}
\put(11,23){\framebox(46,16){{\scriptsize $ (Q'\otimes I_m)\!\!\begin{bmatrix}
P_1(s)&&\\&\ddots&\\&&P_n(s)
\end{bmatrix}\!\!(Q\otimes I_m)$}}}
\put(57,31){\line(1,0){6}}
\put(63,31){\vector(0,-1){22}}
\put(63,8){\circle{2}}
\put(68,8){\vector(-1,0){4}}
\put(62,8){\vector(-1,0){5}}
\put(11,0){\framebox(46,16){{\scriptsize $ (Q'E\otimes I_m)\!\!\begin{bmatrix}\!
W_1(s)\!&&\\&\!\ddots\!&\\&&\!W_l(s)\!
\end{bmatrix}\!\!(E'Q\otimes I_m)$}}}
\put(11,8){\line(-1,0){6}}
\put(5,8){\vector(0,1){22}}
\put(2,28){\makebox(2,3){$-$}}
\end{picture}
\caption{Stability problem corresponds to synchronization with nonuniform communications.}
\label{fig:synchronization-communication}
\end{center}
\end{figure}

We consider the synchronization problem when the edge dynamics model the nonuniform communication environments among the agents. Assume in this subsection that the graph is undirected, i.e., $a_{ij}C_{ij}(s)\!=\!a_{ji}C_{ji}(s)$ for all $i,j\in \mathcal{V}$. In this case, there holds $(\mathbf{1}\!\otimes\! I_m)'\boldL(s)\!=\!0$.
Since $QQ'\!=\!I-\frac{1}{n} \mathbf{1} \mathbf{1}'$, we have
\begin{multline*}
    (Q'\otimes I_m)\boldsymbol{P}(s)\boldL(s)(Q\otimes I_m)\\
    =(Q'\otimes I_m)\boldsymbol{P}(s)(Q\otimes I_m)(Q'\otimes I_m)\boldL(s)(Q\otimes I_m).
\end{multline*}
Further, for an undirected graph, one can assign an arbitrary direction to each edge and get the incidence matrix
\begin{align*}
[E]_{ik}=\begin{cases}1, &\text{if $i$ is the head node of $e_k$},\\-1, &\text{if $i$ is the tail node of $e_k$},\\0,&\text{otherwise}.\end{cases}
\end{align*}
We rename nonzero $a_{ij}C_{ij}(s), i \!>\! j$ as $W_k(s), k \!=\! 1, \dots, l$ and write
$\boldsymbol{W}(s)\!=\!\diag \{W_1(s), W_2(s), \dots, W_l(s)\}$.
Then $\boldL(s)$ has the factorization
\[
\boldL(s) = (E\otimes I_m) \boldsymbol{W}(s) (E\otimes I_m)'.
\]
It follows that
\begin{multline*}
    (Q'\otimes I_m)\boldsymbol{P}(s)\boldL(s)(Q\otimes I_m)\\
    =(Q'\otimes I_m)\boldsymbol{P}(s)(Q\otimes I_m)(Q'E\otimes I_m)\boldsymbol{W}(s)(E'Q\otimes I_m).
\end{multline*}
In view of the above identity, the feedback loop in Figure \ref{synchronization-stability} can be redrawn to the form of Figure \ref{fig:synchronization-communication}, which facilitates the derivation of a synchronization condition presented in the following theorem.

\begin{theorem}
\label{thm:undirected}
Let $\mathbb{G}$ be connected, $\boldsymbol{P}(s)$ be frequency-wise semi-sectorial with residue matrices at poles in $j\Omega$ being sectorial, and $\boldsymbol{W}(s)$ be frequency-wise sectorial. Then synchronization is achieved if
\begin{align*}
\max_i \overline{\phi} (P_i(j\omega)) + \max_k \overline{\phi} (W_k(j\omega))&< \pi,\\
\min_i \underline{\phi} (P_i(j\omega)) + \min_k \underline{\phi} (W_k(j\omega))&> -\pi
\end{align*}
for all $\omega\in[0,\infty]\backslash \Omega$.
\end{theorem}

\begin{pf}
It suffices to show the stability of $\hat{\boldsymbol{P}}(s)\# \hat{\boldL}(s)$, where
\begin{align*}
\hat{\boldsymbol{P}}(s)&=(Q'\otimes I_m)\boldsymbol{P}(s)(Q\otimes I_m)\\
\hat{\boldL}(s)&=(Q'E\otimes I_m)\boldsymbol{W}(s)(E'Q\otimes I_m).
\end{align*}
Clearly, $\hat{\boldsymbol{P}}(s)$ is a compression of $\boldsymbol{P}(s)$. Then, the residue matrix of $\hat{\boldsymbol{P}}(s)$ at $j\omega_k$, given by
$$(Q'\otimes I_m)\diag\{M_{k1},\dots, M_{kn}\}(Q\otimes I_m),$$ is sectorial and thus nonsingular for all $k=0,\dots,q$. Note that when the graph is connected, $\mathrm{kernel}(E')=\mathrm{span}\{\mathbf{1}_n\}$ and thus $\mathrm{kernel}(E')\perp\mathrm{range}(Q)$. Therefore, $E'Q$ has full rank and $\hat{\boldL}(s)$ is a compression of $\boldsymbol{W}(s)$.
Hence, $\hat{L}(j\omega_k)$ is nonsingular for all $k=0,\dots, q$ and there is no unstable pole zero cancellation between $\hat{\boldP}(s)$ and $\hat{\boldL}(s)$.
By Lemma~\ref{lemma:semi-compression}, we have
\begin{align*}
\overline\phi(\hat{\boldsymbol{P}}(j\omega))&\leq \overline\phi(\boldsymbol{P}(j\omega))=\max_i \overline\phi(P_i(j\omega)),\\
\underline\phi(\hat{\boldsymbol{P}}(j\omega))&\geq \underline\phi(\boldsymbol{P}(j\omega))=\min_i \underline\phi(P_i(j\omega))
\end{align*}
for all $\omega\in[0,\infty]\backslash\Omega$, and
\begin{align*}
\overline\phi(\hat{\boldL}(j\omega))&\leq \overline{\phi}(\boldsymbol{W}(j\omega))=\max_k \overline{\phi}(W_k(j\omega)),\\
\underline\phi(\hat{\boldL}(j\omega))&\geq \underline{\phi}(\boldsymbol{W}(j\omega))=\min_k \underline{\phi}(W_k(j\omega))
\end{align*}
for all $\omega\in[0,\infty]$.

When the phase conditions in the theorem hold, we have
\begin{align*}
\overline{\phi}(\hat{\boldsymbol{P}}(j\omega))\! +\! \overline{\phi}(\hat{\boldL}(j\omega)) \!<\!\pi,\ \
\underline{\phi}(\hat{\boldsymbol{P}}(j\omega)) \!+\! \underline{\phi}(\hat{\boldL}(j\omega)) \!>\!-\pi
\end{align*}
for all $\omega\in[0,\infty]\backslash\Omega$.
It then follows from Lemma~\ref{thm:gspt} that $\hat{\boldsymbol{P}}(s)\# \hat{\boldL}(s)$ is stable, which completes the proof. \hfill\qed
\end{pf}

\begin{remark}
Requiring that $\boldsymbol{P}(s)$ be frequency-wise semi-sectorial in fact requires that $P_1(s),\dots,P_n(s)$ be jointly frequency-wise semi-sectorial, i.e., for each $\omega\in[0,\infty]$, $W(P_i(j\omega)),i=1,\dots,n$, are simultaneously contained in a closed half plane. Similarly for $\boldsymbol{W}(s)$.
\end{remark}

Theorem \ref{thm:undirected} guarantees the output synchronization of the  heterogeneous agents over nonuniform edge dynamics by imposing only local phase conditions. These conditions are independent of the network topology. Such a result would generalize positive real type conditions \cite{Allgower2014}. There is no requirement that $P_i(j\omega)$ be positive real across all frequencies; they may have phases within $[-\pi/2,\pi/2]$ at some frequencies and beyond $[-\pi/2,\pi/2]$ at others.

Note that the conditions in Theorem \ref{thm:undirected} scale well with the size of the network. In particular, when a new agent joins the network or a new communication link is established, the information about the new entry simply needs
to be compared with the outcome of the phase analysis previously conducted for the
original network with $n$ agents and $l$ links. In other words, re-performing a centralized
phase analysis involving all nodes is not necessary. These conditions are considered as having a ``plug and play'' property.

\subsection{Synchronization with agent-dependent controllers}
In this subsection, we consider a different scenario when the edge dynamics model the controllers applied to the agents. We assume that all the edges connecting to an agent have the same dynamic part, i.e., $C_{ij}(s)=C_i(s)$ for all $j$ with $(j,i)\in\mathcal{E}$, where $C_i(s)\in\mathcal{RH}_{\infty}^{m\times m}$ has full rank at $j\Omega$ for all $i$. This means that each agent is controlled by controllers with the same dynamic part regardless the source of the feedback signal. Agent-dependent controllers were also considered in other studies of synchronization, for instance \cite{SuHuang}.



Denote by $L_0$ the Laplacian matrix associated with the underlying directed network where the edge weights are given by the static gains $a_{ij}$. Then, the block diagram in Figure \ref{network_framework} can be redrawn to the form of Figure \ref{fig: partialsame}.

\setlength{\unitlength}{1.23mm}
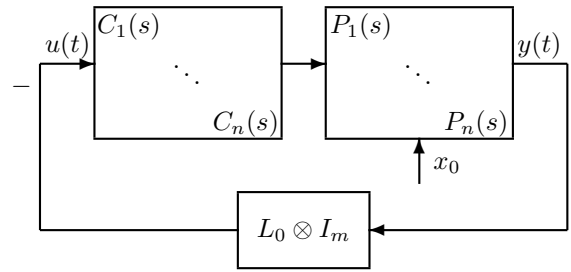
\begin{figure}[htb]
\begin{center}
\begin{picture}(60,29)
\thicklines
\put(2,22){\vector(1,0){6}}
\put(8,14){\framebox(20,14){$\displaystyle \begin{array}{ccc}
                                                    C_1(s) & & \\
                                                     & \ddots & \\
                                                     & & C_n(s) \end{array}$}}
\put(28,22){\vector(1,0){5}}
\put(33,14){\framebox(20,14){$\displaystyle \begin{array}{ccc}
                                                    P_1(s) & & \\
                                                     & \ddots & \\
                                                     & & P_n(s) \end{array}$}}
\put(43,9){\vector(0,1){5}}
\put(44,9){\makebox(4,4){$x_0$}}
\put(53,22){\line(1,0){6}}
\put(59,22){\line(0,-1){18}}
\put(59,4){\vector(-1,0){21.5}}
\put(23.5,0){\framebox(14,8){$L_0\otimes I_m$}}
\put(23.5,4){\line(-1,0){21.5}}
\put(2,4){\line(0,1){18}}
\put(3,22){\makebox(4,4){$\displaystyle u(t)$}}
\put(54,22){\makebox(4,4){$\displaystyle y(t)$}}
\put(-1,18){\makebox(2,3){$-$}}
\end{picture}
\vspace{-5pt}
\caption{Synchronization with agent-dependent controllers.}
\label{fig: partialsame}
\end{center}
\end{figure}

Suppose the graph has $n_1$ roots and $\nu$ strongly connected components. One can relabel the nodes to form $\nu$ groups as in (\ref{strongcd})
so that the nodes in each group correspond to a strongly connected component and the first component contains all the roots. The Laplacian $L_0$ is then of the form (\ref{eq: L_decomp}).
Let
\begin{equation}
\boldsymbol{P}_{\kappa\kappa}(s)=\diag \{P_{n_{\kappa-1}+1}(s), \dots, P_{n_{\kappa}}(s)\} \label{eq: agents_decomp}
\end{equation}
and the controllers be grouped accordingly as
\begin{equation}
\boldsymbol{C}_{\kappa\kappa}(s)=\diag \{C_{n_{\kappa-1}+1}(s), \dots, C_{n_{\kappa\kappa}}(s)\},
\label{eq: controller_decomp}
\end{equation}
where $\kappa=1,\dots,\nu$, $n_0=0$.
We have the following synchronization result.

\begin{theorem}\label{thm:syn_agent_alys}
If $P_{i}(s)C_{i}(s)$ are jointly frequency-wise semi-sectorial for $i=n_{\kappa-1}+1,\dots, n_{\kappa}$ and
\begin{multline*}
\mathop{\bigcup}\limits_{n_{\kappa-1}+1\leq i \leq n_\kappa} \Phi \left( P_i(j\omega) C_{i}(j\omega)\right) \\ \subset (-\pi+\phi_{\ess}(L_{\kappa\kappa}), \pi-\phi_{\ess}(L_{\kappa\kappa}))
\end{multline*}
holds for all $\kappa=1,\dots, \nu$ and $\omega\in[0,\infty]\backslash \Omega$, then synchronization is achieved.
\end{theorem}

\begin{pf}
Denote
\begin{align*}
u_{\kappa\kappa}(t)&=\begin{bmatrix}
u_{n_{\kappa-1}+1}(t)' &\cdots & u_{n_\kappa}(t)'
\end{bmatrix}',\\
y_{\kappa\kappa}(t)&=\begin{bmatrix}
y_{n_{\kappa-1}+1}(t)' &\cdots & y_{n_\kappa}(t)'
\end{bmatrix}',\\
x_{0(\kappa\kappa)}&=\begin{bmatrix}
x_{0(n_{\kappa-1}+1)}' &\cdots & x_{0n_\kappa}'
\end{bmatrix}',\\
\boldG_{\kappa\kappa}(s)&=\boldP_{\kappa\kappa}(s)\boldC_{\kappa\kappa}(s),
\end{align*}
where $\kappa=1,\dots, \nu$. From the agents' dynamics \eqref{response} we have
\begin{align}\label{ykk}
    y_{\kappa\kappa}(s)=\boldG_{\kappa\kappa}(s)u_{\kappa\kappa}(s)+\boldsymbol{H}_{\kappa\kappa}(s)x_{0(\kappa\kappa)},
\end{align}
where
\begin{multline*}
\boldsymbol{H}_{\kappa\kappa}(s)=\diag\{C_{n_{\kappa-1}+1}, \dots, C_{n_k}\}\\
\cdot (sI-\diag\{A_{n_{\kappa-1}+1}, \dots, A_{n_k}\})^{-1}.
\end{multline*}
From the synchronization protocol \eqref{procompact}, we have
\begin{align*}
    u_{\kappa\kappa}(s)\!=\!-(L_{\kappa\kappa}\!\otimes\! I_m)y_{\kappa\kappa}(s)\!-\!\sum_{l=1}^{\kappa-1} (L_{\kappa l}\otimes I_m) y_{ll}(s),
\end{align*}
which can be substituted into \eqref{ykk} so as to yield
\begin{align}\label{y11tf}
    y_{11}(s)=     {\big(}I+\boldsymbol{G}_{11}(s)(L_{11}\!\otimes\! I_m){\big)}^{-1}\boldsymbol{H}_{11}(s)x_{0(11)}
\end{align}
and
\begin{align}
    &y_{\kappa\kappa}(s)\!=\!{\big(}I+\boldG_{\kappa\kappa}(s)(L_{\kappa\kappa}\!\otimes\! I_m){\big)}^{-1}\nonumber\\
    &\left(\! \boldsymbol{H}_{\kappa\kappa}(s)x_{0(\kappa\kappa)}\!-\! \boldG_{\kappa\kappa}(s)\!
    \left( \sum_{l=1}^{\kappa-1} (L_{\kappa l}\!\otimes\! I_m) y_{ll}(s)\! \right)\!\right), \label{ykktf}
\end{align}
$\kappa=2,\dots,\nu$. One can see that the dynamics of the root agents are independent
of those of the remaining agents.

We first show that all the roots reach synchronization. In view of \eqref{y11tf}, by using the same analysis as in Section~\ref{setup}, we know that the roots reach synchronization if and only if the feedback system
\begin{align}
(Q'_1\otimes I_m)\boldsymbol{G}_{11}(s)\# (L_{11}\otimes I_m)(Q_1\otimes I_m)\label{gof1}
\end{align}
is stable, where $Q_1$ is such that $\begin{bmatrix}\frac{\mathbf{1}_{n_1}}{\sqrt{n_1}}&Q_1\end{bmatrix}$ is an orthogonal matrix. Recall that $L_{11}$ is the Laplacian of the strongly connected component consisting of all the roots. It has a positive left eigenvector $v_1$ corresponding to the zero eigenvalue, i.e., $v'_1L\!=\!0$. Let $V_1\!=\!\mathrm{diag}\{v_1\}$. Then $V_1L_{11}$ has $\mathbf{1}$ being a common left and right eigenvector corresponding to the zero eigenvalue.
We further denote
\begin{align}
\hat{\boldsymbol{P}}(s)&=(Q'_1\!\otimes\! I_m)\boldsymbol{G}_{11}(s)(V_1^{-1}Q_1\otimes I_m),\label{Phat}\\
\hat{L}&=(Q'_1V_1L_{11}Q_1)\otimes I_m.\label{Lhat}
\end{align}
Then,
\begin{align*}
\hat{\boldsymbol{P}}(s)\hat{L}
&\!=\!(Q'_1\!\otimes\! I_m)\boldsymbol{G}_{11}(s)(V_1^{-1}Q_1Q_1'V_1L_{11}Q_1\!\otimes\! I_m)\\
&\!=\!(Q'_1\!\otimes\! I_m)\boldsymbol{G}_{11}(s) \!\!\left(\!V_1^{-1} \!\!\left(\!I\!-\!\frac{\mathbf{1}\mathbf{1}'}{n_1}\right)\!\! V_1L_{11}Q_1\!\otimes\! I_m \!\!\right)\\
&\!=\! (Q'_1\!\otimes\! I_m)\boldsymbol{G}_{11}(s)(L_{11}\!\otimes\! I_m)(Q_1\!\otimes\! I_m),
\end{align*}
where the last equality is due to the fact $\mathbf{1}'V_1L_{11}=0$. This implies that the stability of \eqref{gof1} is in turn equivalent to the stability of $\hat{\boldsymbol{P}}(s)\#\hat{L}$.

Note that $\hat{\boldsymbol{P}}(s)$ is a compression of $\boldsymbol{G}_{11}(s)(V_1^{-1}\otimes I_m)$ and $\hat{L}$ is a compression of $V_1L_{11}\otimes I_m$.
By Lemma~\ref{lemma:semi-compression},
\begin{align*}
\overline\phi(\hat{\boldsymbol{P}}(j\omega))\!\leq\! \overline\phi(\boldsymbol{G}_{11}(j\omega)),\ \
\underline\phi(\hat{\boldsymbol{P}}(j\omega))\!\geq\! \underline\phi(\boldsymbol{G}_{11}(j\omega))
\end{align*}
for all $\omega\in[0,\infty]\backslash\Omega$, and
\begin{align*}
\overline\phi(\hat{L})&\leq \overline\phi((V_1L_{11})\otimes I_m)= \phi_{\ess}(L_{11}),\\
\underline\phi(\hat{L})&\geq \underline\phi((V_1L_{11})\otimes I_m)= -\phi_{\ess}(L_{11}),
\end{align*}
where the equalities follow from Lemma \ref{essphaseLap}.
In view of the condition in the theorem, there holds
\begin{align*}
\overline\phi(\hat{\boldsymbol{P}}(j\omega)) +  \overline\phi(\hat{L}) <\pi,\ \
\underline\phi(\hat{\boldsymbol{P}}(j\omega)) +\underline\phi(\hat{L}) >-\pi
\end{align*}
for all $\omega\!\in\![0,\infty]\backslash\Omega$, yielding that $\hat{\boldsymbol{P}}(s)\#\hat{L}$ is stable. This shows that the roots reach synchronization. In view of the analysis in Section 2, the roots converge to a common trajectory determined by the modes in $j\Omega$ and their initial conditions.

Next we show that the agents in the second group converge to the same synchronizing trajectory of the root agents. Let us decompose $y_{11}(t)$ into two parts:
\[
y_{11}(t)=\mathbf{1}_{n_1}\otimes y_{11}^{\ave}(t)+y_{11}^{\dis}(t),
\]
where $y_{11}^{\ave}(t)=\frac{1}{n_1}(\mathbf{1}_{n_1}\otimes I_m)'y_{11}(t)$. In view of \eqref{ykktf}, we have
\begin{align*}
y_{22}(s)&\!=\!  {\big(}I\!+\!\boldsymbol{G}_{22}(s)(L_{22}\!\otimes\! I_m){\big)}^{-1} \!{\Big(}\!  \boldsymbol{H}_{22}(s)x_{0(22)}\!-\!\boldsymbol{G}_{22}(s) \\
& \cdot (L_{21}\mathbf{1}_{n_1}\otimes y_{11}^{\ave}(s)) -\boldsymbol{G}_{22}(s)(L_{21}\otimes I_m)y_{11}^{\dis}(s) {\Big)}\\
&\!=\! {\big(}I\!+\!\boldsymbol{G}_{22}(s)(L_{22}\!\otimes\! I_m){\big)}^{-1} {\Big(} \boldsymbol{H}_{22}(s)x_{0(22)}  \\
& +\boldsymbol{G}_{22}(s) (L_{22}\otimes I_m)(\mathbf{1}_{n_2-n_1}\otimes y_{11}^{\ave}(s)) \\
& -\boldsymbol{G}_{22}(s)(L_{21}\otimes I_m) y_{11}^{\dis}(s) {\Big)},
\end{align*}
where the last equality is due to $L_{21}\mathbf{1}_{n_1}=-L_{22}\mathbf{1}_{n_2-n_1}$.
The dynamics of the agents $\{n_1+1,\dots, n_2\}$ are then represented by a tracking block diagram in Figure \ref{fig:y22}. One can see that if
\begin{align}
\boldsymbol{G}_{22}(s)\# (L_{22}\otimes I_m)\label{gof2}
\end{align}
is stable, then the steady-state response of $y_{22}(t)$ to $y_{11}^{\dis}(t)$ and $x_{0(22)}$ are zero. The transfer function from $\mathbf{1}_{n_2-n_1}\otimes y_{11}^{\ave}(t)$ to $y_{22}(t)$ is given by
\begin{equation*}
\label{eq:sensitivity}
(I+\boldG_{22}(s)(L_{22}\otimes I_m))^{-1}\boldG_{22}(s)(L_{22}\otimes I_m),
\end{equation*}
which evaluated at $j\omega_k$ equals to identity for all $k=0,\dots,q$. Therefore, the steady-state of $y_{22}(t)$ equals the steady-state of $\mathbf{1}_{n_2-n_1}\otimes y_{11}^{\ave}(t)$, which means that the agents in the second group will converge to the same synchronizing trajectory as the root agents.

\begin{figure}[htb]
\begin{center}
\begin{picture}(64,23)\scriptsize
\thicklines
\put(0,17){\vector(1,0){16}}
\put(-1,17){\makebox(16,4){$\displaystyle \mathbf{1}_{n_2\!-\!n_1}\!\otimes\! y_{11}^{\ave}(t)$}}
\put(16,14){\framebox(12,6){$\displaystyle L_{22}\otimes I_m$}}
\put(28,17){\vector(1,0){5}}
\put(8,6){\vector(1,0){8}}
\put(10,6){\makebox(4,4){$\displaystyle y_{11}^{\dis}(t)$}}
\put(16,3){\framebox(12,6){$\displaystyle -L_{21}\!\otimes\! I_m$}}
\put(28,6){\line(1,0){2}}
\put(30,6){\vector(1,3){3.4}}
\put(34,17){\circle{2}}
\put(35,17){\vector(1,0){4}}
\put(39,14){\framebox(16,6){$\displaystyle \boldsymbol{G}_{22}(s)$}}
\put(47,9){\vector(0,1){5}}
\put(49,9){\makebox(4,4){$x_{0(22)}$}}
\put(55,17){\vector(1,0){8}}
\put(56,17){\makebox(6,4){$\displaystyle y_{22}(t)$}}
\put(60,17){\line(0,-1){14}}
\put(60,3){\vector(-1,0){5}}
\put(39,0){\framebox(16,6){$\displaystyle L_{22}\otimes I_m$}}
\put(39,3){\line(-1,0){5}}
\put(34,3){\vector(0,1){13}}
\put(34.5,11.5){\makebox(2,3){$-$}}
\end{picture}
\vspace{-3pt}
\caption{Tracking block diagram for the second group.}
\label{fig:y22}
\end{center}
\end{figure}
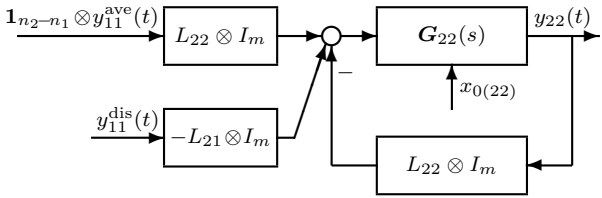

Now we show under the stated conditions, (\ref{gof2}) is indeed stable. By Lemma \ref{epLii}, there exists a positive diagonal $V_2$ such that $$\phi_{\ess}(L_{22})=\overline{\phi}(V_2 L_{22}).$$ Then the condition in the theorem implies that
\begin{align*}
    &\overline{\phi}(\boldsymbol{G}_{22}(j\omega)(V_2^{-1}\!\otimes\! I_m))+\overline{\phi}(V_2 L_{22}\!\otimes\! I_m)<\pi,\\
    &\underline{\phi}(\boldsymbol{G}_{22}(j\omega)(V_2^{-1}\!\otimes\! I_m))+\underline{\phi}(V_2 L_{22}\!\otimes\! I_m)>-\pi.
\end{align*}
By Lemma \ref{thm:gspt}, the feedback system $$\boldsymbol{G}_{22}(s)(V_2^{-1}\otimes I_m)\# (V_2 L_{22}\otimes I_m)$$ is stable, which is equivalent to (\ref{gof2}) being stable.

Synchronization of the remaining groups of the agents can be shown similarly. The proof is completed. \hfill\qed
\end{pf}

One can see from the proof that the synchronization
problem of a dynamical network which has a spanning
tree can be decomposed into two subproblems: synchronization of the roots and sequential tracking of the remaining agents.
Moreover, the synchronized trajectory depends only on the initial values of the roots while the other agents track this synchronized trajectory.

Note that $\phi_{\mathrm{ess}}(L_{\kappa\kappa}),\kappa=1,\dots,\nu$ are completely determined by the network topology and do not change over frequency. Conditions in Theorem~\ref{thm:syn_agent_alys} reflect the trade-off between network and dynamics. Since $\phi_{\mathrm{ess}}(L_{\kappa\kappa})<\pi/2$ for all $\kappa=1,\dots, \nu$, this theorem indicates that if $\boldsymbol{P}\boldsymbol{C}$ is positive real, then synchronization is reached. In another case where the network is undirected, the theorem implies that synchronization is reached if $\boldsymbol{P}\boldsymbol{C}$ is frequency-wise semi-sectorial with $[\underline\phi (\boldsymbol{P}\boldsymbol{C}), \overline\phi  (\boldsymbol{P}\boldsymbol{C})] \subset (-\pi,\pi)$.

\section{Controller Design for Synchronization}\label{sec: design}
In this section, we study the synthesis problem, i.e., design controllers so as to achieve synchronization. We seek answers to two main issues. The first one is to characterize the synchronizability condition with respect to the diversity of the agents. We will see that phase bounded cones, in comparison with norm bounded balls, give a more natural fit in characterizing the allowable agent diversities. The second issue is to give a design procedure which produces synchronizing controllers, when the synchronizability condition is satisfied.

For the sake of brevity, a major technical lemma is given in the Appendix. It plays critical roles in establishing the synchronizability conditions.

Before proceeding, we provide a useful interpolation algorithm by using the Lagrange polynomial \cite{Lagrange-interpolation}. It gives a transfer function matrix $C(s)\!\in\!\mathcal{RH}_{\infty}^{m\times m}$ satisfying the interpolation condition on $j\Omega$
\begin{align*}
C(0)&=K_0\in\mathbb{R}^{m\times m} \text{ and }\\
C(j\omega_k)&=K_k\in\mathbb{C}^{m\times m}, k=1,\dots,q.
\end{align*}

\begin{algorithm}[htb]
\caption{}
\label{al:agent-dependent}
\begin{algorithmic}[1]
\State Let $z_0\!=\!1$, $\displaystyle z_k\!=\!\frac{1\!-\!j\omega_k}{1\!+\!j\omega_k}$,  $\displaystyle z_{q+k}\!=\!\frac{1\!+\!j\omega_k}{1\!-\!j\omega_k}$, $k\!=\!1,\dots, q$.
\label{step1}
\State Let
\begin{align*}
f(z^{-1})&=K_{0}\prod_{l=1}^{2q}\frac{z^{-1}-z_l}{z_0-z_l}
+ \sum_{k=1}^q K_k \prod_{l=0, l\neq k}^{2q} \frac{z^{-1}-z_l}{z_k-z_l}\\
&+ \sum_{k=1}^q \bar{K}_k \prod_{l=0, l\neq q+k}^{2q} \frac{z^{-1}-z_l}{z_{q+k}-z_l}.
\end{align*}
\State Letting $z^{-1}=\frac{1-s}{1+s}$ in $f(z^{-1})$ gives $C(s)$.
\end{algorithmic}
\end{algorithm}

\subsection{Synchronization with agent-dependent controllers}
Let the agent dynamics be given in the form of (\ref{agent}), where the residue matrices $M_{ki},k=0,1,\dots,q,i=1,2,\dots,n$ are assumed to be nonsingular. We consider the case of agent-dependent controllers as in Figure \ref{fig: partialsame}. The synchronization protocol then becomes
\begin{align}\label{protocal}
u_i(s)&=\sum_{(j,i)\in\mathcal{E}} a_{ij}C_i(s)(y_j(s)-y_i(s)).
\end{align}
While $C_1(s),\dots,C_n(s)$ are given {\em a priori} in the analysis problem, they are to be designed in the synthesis problem.

As before, we divide the agents into groups of the form \eqref{eq: agents_decomp} and the agent-dependent controllers into groups of the form (\ref{eq: controller_decomp}) accordingly.
Denote
\begin{align}
\boldsymbol{M}_{k\kappa}=\diag\{M_{k(n_{\kappa-1}+1)},\dots, M_{kn_{\kappa}}\},\label{Mkk}
\end{align}
where $\kappa=1,\dots,\nu,k=0,\dots,q$ and $n_0=0$.

\begin{dproblem}\label{pro1}
Design controllers $C_i(s)$ so that the agents of the form \eqref{agent} achieve synchronization under protocol (\ref{protocal}).
\end{dproblem}

Having each agent equipped with its own controller gives a large space of design freedoms. If the controllers are designed judiciously in accordance with the features of the individual agents, the heterogeneity of the agents may get smoothed out by the controllers and consequently its effect on synchronizability may be significantly reduced or even eliminated. It turns out this is indeed the case.

\begin{theorem}\label{thm:syn_agent_design}
Design Problem \ref{pro1} is always solvable.
\end{theorem}

\begin{pf}
We design $C_i(s)\!=\!\epsilon\tilde{C}_i(s)$, where $\tilde{C}_i(s)$ is obtained by interpolation via Algorithm 1 such that
\begin{align*}
\tilde{C}_i(0)=M^{-1}_{0i} \text{ and } \tilde{C}_i(j\omega_k)=M^{-1}_{ki},k=1,\dots,q.
\end{align*}
Let
    $\boldsymbol{K}_{k\kappa}\!=\!\diag\{M^{-1}_{k(n_{\kappa-1}+1)},\dots, M^{-1}_{kn_{\kappa}}\}$ for $\kappa\!=\!1,\dots,\nu$, $k\!=\!0,\dots,q$.
Then
\begin{align*}
\angle \lambda (\boldsymbol{M}_{k\kappa}\boldsymbol{K}_{k\kappa}(L_{\kappa\kappa}\!\otimes\! I_m))\!=\!\angle \lambda(L_{\kappa\kappa}\otimes I_m)\!\in\! (-\pi/2,\pi/2)
\end{align*}
for $\kappa\!=\!1,\dots,\nu,k\!=\!0,\dots,q$. By Lemma \ref{mtl}, synchronization is reached with $C_i(s)$ for sufficiently small $\epsilon\!\in\!(0,\epsilon^*)$, where $\epsilon^*$ can be estimated from the given data.\hfill\qed
\end{pf}

The constructive proof gives a controller design method. The design confirms our previous speculation: The controllers ``homogenize'' the diversely different agents in a way that they make the residues of $P_i(s)C_i(s)$ at the imaginary-axis poles all equal to identity (up to a scalar multiplication).

We mention that in the case of consensus problem, $P_i(s)$ share only one common pole on the imaginary axis, i.e.,
\begin{align}\label{agent:consensus}
P_i(s)=\frac{M_{0i}}{s}+\Delta_i(s).
\end{align}
By Theorem \ref{thm:syn_agent_design}, the consensus problem is always solvable with agent-dependent controllers. What is more, static controllers $C_i(s)=\epsilon M_{0i}^{-1}, i=1,\dots, n$ for $\epsilon\!\in\!(0,\epsilon^*)$ will make the agents achieve consensus, where $\epsilon^*$ can be estimated from the given data.

\subsection{Synchronization with a uniform controller}\label{synuniform}

We have seen previously that the use of agent-based controllers facilitates the synchronization design in the sense that it makes the problem always solvable. Nevertheless, such design does not scale with the size of the network.

\begin{figure}[htb]
\begin{center}
\begin{picture}(53,29)
\thicklines
\put(4,20){\vector(1,0){8}}
\put(12,12){\framebox(32,16){$\displaystyle \begin{array}{ccc}
                                                    P_1(s)C(s) & & \\
                                                     & \ddots & \\
                                                     & & P_n(s)C(s) \end{array}$}}
\put(28,8){\vector(0,1){4}}
\put(29,8){\makebox(4,4){$\displaystyle x_0$}}
\put(44,20){\line(1,0){8}}
\put(52,20){\line(0,-1){16}}
\put(52,4){\vector(-1,0){17}}
\put(21,1){\framebox(14,6){$L_0\otimes I_m$}}
\put(21,4){\line(-1,0){17}}
\put(4,4){\line(0,1){16}}
\put(6,20){\makebox(4,4){$\displaystyle u(t)$}}
\put(46,20){\makebox(4,4){$\displaystyle y(t)$}}
\put(0.5,15){\makebox(2,3){$-$}}
\end{picture}
\vspace{-5pt}
\caption{Synchronization with a uniform controller.}
\label{fig: directed}
\end{center}
\end{figure}
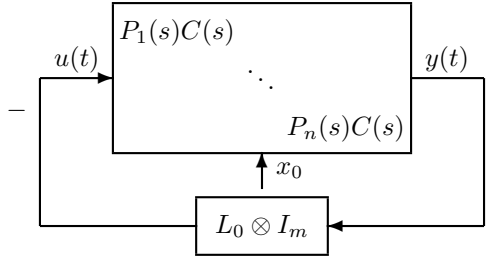

In the face of a large-scale network, scalability becomes a very important issue. In this respect, it would be more appealing if one can design a uniform controller for all of the agents, i.e., $C_i(s)=C(s),i=1,\dots,n$, to achieve synchronization. The synchronization protocol then becomes
\begin{align}\label{protocal_uniform}
u_i(s)&=\sum_{(j,i)\in\mathcal{E}} a_{ij}C(s)(y_j(s)-y_i(s))
\end{align}
and the block diagram in Figure \ref{fig: partialsame} can be redrawn to the form of Figure \ref{fig: directed}.

Restricting controllers to be uniform poses a great challenge to the design. Recall that synchronization is essentially a feedback stability problem on the disagreement subspace. In this aspect, using a uniform controller to synchronize a large group of heterogeneous agents is in essence a simultaneous stabilization problem. We wish to characterize the solvability condition and also find an algorithm that generates such a uniform controller if the problem is solvable.


\begin{dproblem}\label{pro:3}
Design a uniform controller $C(s)$ so that the agents of the form \eqref{agent} achieve synchronization under protocol (\ref{protocal_uniform}).
\end{dproblem}


The following theorem provides an answer to solvability. The allowable diversity of agents is characterized explicitly in terms of a phase condition.

\begin{theorem}\label{thm:synbility}
Design Problem \ref{pro:3} is solvable if there exist nonsingular $K_0\!\in\!\mathbb{R}^{m\times m}$ and $K_1, \dots, K_q\!\in\!\mathbb{C}^{m\times m}$ such that
\begin{multline*}
\Phi(\boldsymbol{M}_{k\kappa}(I_{n_{\kappa}-n_{\kappa-1}}\!\otimes\! K_k))\\
\!\subset\! \left( -\pi/2\!+\!\phi_{\ess}(L_{\kappa\kappa}), \pi/2\!-\!\phi_{\ess}(L_{\kappa\kappa}) \right)
\end{multline*}
holds for $k=0,\dots, q,\kappa=1,\dots,\nu$.
\end{theorem}

\begin{pf}
We design $C(s)\!=\!\epsilon\tilde{C}(s)$, where $\tilde{C}(s)$ is obtained by interpolation via Algorithm 1 such that
\begin{align*}
\tilde{C}(0)=K_0 \text{ and } \tilde{C}(j\omega_k)=K_k,k=1,\dots,q.
\end{align*}
We shall show such a controller renders synchronization when $\epsilon>0$ is sufficiently small.

By Lemmas \ref{essphaseLap} and \ref{epLii}, there are positive diagonal matrices $V_{\kappa}$ such that $\phi_{\ess}(L_{\kappa\kappa})\!=\!\overline{\phi}(V_{\kappa}L_{\kappa\kappa}),\kappa\!=\!1,\dots,\nu$.
Denote $\boldsymbol{K}_{k\kappa}\!=\!I_{n_{\kappa}-n_{\kappa-1}}\!\otimes\! K_k$, $\kappa\!=\!1,\dots,\nu$, $k\!=\!0,\dots,q$. Then
\begin{equation*}
\boldsymbol{M}_{k\kappa}\boldsymbol{K}_{k\kappa}(L_{\kappa\kappa}\!\otimes\! I_m)\!=\!\boldsymbol{M}_{k\kappa}\boldsymbol{K}_{k\kappa}(V^{-1}_{\kappa}\otimes I_m)(V_{\kappa}L_{\kappa\kappa}\otimes I_m).
\end{equation*}
Applying Lemma \ref{lemma:semi-major} yields
\begin{multline*}
\angle \lambda (\boldsymbol{M}_{k\kappa}\boldsymbol{K}_{k\kappa}(L_{\kappa\kappa}\!\otimes\! I_m))\\
\in \Phi(\boldsymbol{M}_{k\kappa}\boldsymbol{K}_{k\kappa}(V^{-1}_{\kappa}\otimes I_m))+\Phi(V_{\kappa}L_{\kappa\kappa}\otimes I_m).
\end{multline*}
This, together with the condition stated in the theorem, further yields
$
    \angle \lambda (\boldsymbol{M}_{k\kappa}\boldsymbol{K}_{k\kappa}(L_{\kappa\kappa}\!\otimes\! I_m))\in (-\pi/2,\pi/2).
$
Then Lemma \ref{mtl} tells that synchronization is reached with $C(s)$ for sufficiently small $\epsilon\!\in\!(0,\epsilon^*)$, where $\epsilon^*$ can be estimated from the given data.\hfill\qed
\end{pf}

It is seen that the solvability condition depends only on the phase information of the residues at the imaginary-axis poles of each agent, not at all the gain information. This indicates that the synchronization problem is
likely solvable if the agents have vastly different sizes but similar shapes. Examples of
such agents with large difference in sizes but similarity in shapes are a collection of tigers
and cats, and a collection of large UAVs and small UAVs. This solvability condition is
much weaker than requiring that $P_i(s)$ are all positive real, as indicated in some recent
studies \cite{Allgower2014}. It also gives a clear trade-off between the agent data and the network data. In the special case of an undirected network, $\phi_{\mathrm{ess}}(L_{\kappa\kappa})\!=\!0$. Then the problem is solvable if there exist $K_k$ such that $M_{ki} K_k$ is strictly accretive for all $k=0,\dots, q, i=1,\dots, n$.

The proof is constructive which gives a controller design algorithm once the desired $K_k, k=0,\dots,q$ are found. Finding these $K_k$ is in turn equivalent to solving a group of LMIs:
\begin{equation}\label{eq:LMI_uniform}
\begin{split}
\mathrm{Re} \;e^{j\theta_{\kappa}}\boldsymbol{M}_{k\kappa}(I_{n_{\kappa}-n_{\kappa-1}}\!\otimes\! K_k)&>0,\\
\mathrm{Re} \; e^{-j\theta_{\kappa}}\boldsymbol{M}_{k\kappa}(I_{n_{\kappa}-n_{\kappa-1}}\!\otimes\! K_k)&>0,\kappa=1,\dots,\nu,
\end{split}
\end{equation}
where $\mathrm{Re}$ represents the Hermitian part of a matrix and $\theta_{\kappa}=\phi_{\mathrm{ess}}(L_{\kappa\kappa}), \kappa=1,\dots, \nu$.


The design of synchronizing controllers suggests the use of low gain controllers,
indicating that the coordination among the agents does not need strong action. Instead
it is more critical to have the right directions of the action.


When specializing the results to the consensus of agents given by \eqref{agent:consensus} under consensus protocol \eqref{protocal_uniform}, we have the following corollary of Theorem \ref{thm:synbility}.
\begin{corollary}
The consensus problem is solvable if there exists a nonsingular real matrix $K_0$ such that
\begin{multline*}
\Phi(\boldsymbol{M}_{0\kappa}(I_{n_{\kappa}-n_{\kappa-1}}\!\otimes\!K_0))\\
\!\subset\! \left( -\pi/2\!+\!\phi_{\ess}(L_{\kappa\kappa}), \pi/2\!-\!\phi_{\ess}(L_{\kappa\kappa}) \right)
\end{multline*}
holds for $\kappa=1,\dots, \nu$.
\end{corollary}

Again, checking whether the condition is satisfied is an LMI feasibility problem. When the condition is satisfied, a solution of the problem is given by a static controller $C(s)\!=\!\epsilon K_0$ for $\epsilon\!\in\!(0,\epsilon^*)$, where $\epsilon^*$ can be estimated from the given data.

\section{Simulations}
\label{sec: simulation}

\begin{figure}[htb]
    \centering
    \includegraphics[scale=0.45]{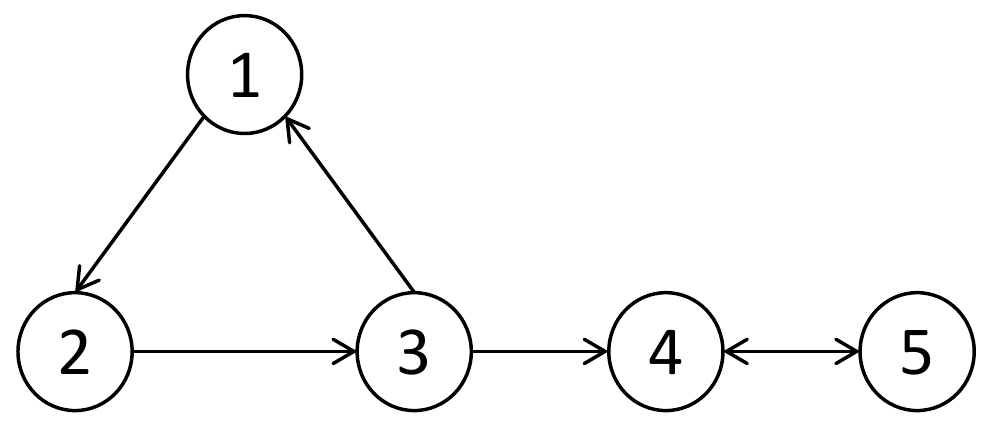}
    \caption{An example network.}
    \label{fig:syn_design}
\end{figure}

\begin{figure*}
    \centering
    \includegraphics[scale=0.65]{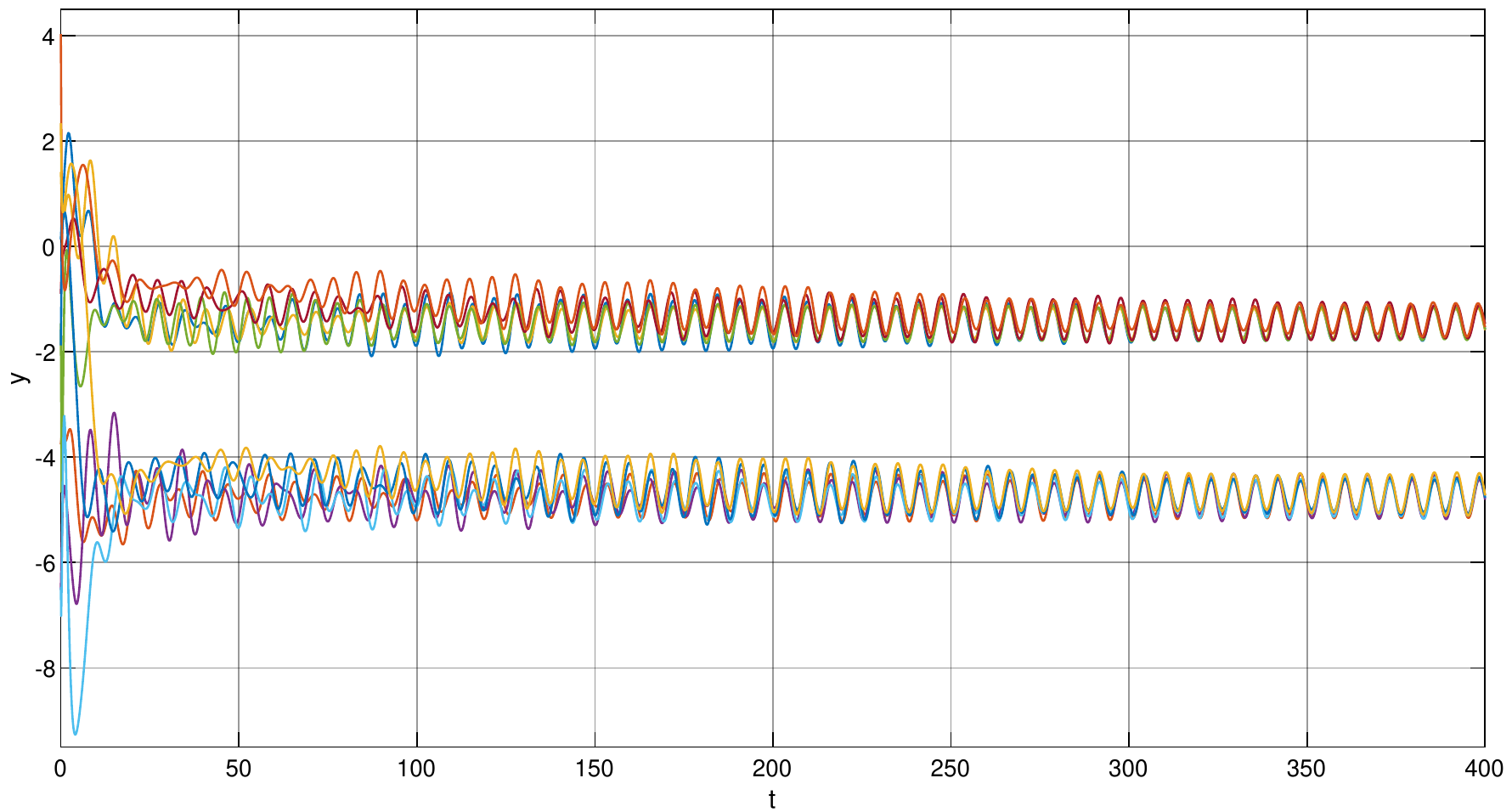}
    \caption{Outputs of the agents.}
    \label{fig:syn_uniform}
\end{figure*}

In this section, we use an example to illustrate the synthesis result Theorem \ref{thm:synbility}. Consider a network consisting of five agents. The agent dynamics are given by
\scriptsize
\begin{equation*}
\begin{split}
P_1(s)&=\frac{\begin{bmatrix}14&2\\5&12\end{bmatrix}}{s} + \frac{\begin{bmatrix}8s-10&12s-2\\14s-6&2s-2\end{bmatrix}}{s^2+1} + \frac{3\begin{bmatrix} s+1&s+4\\s-1&s+3 \end{bmatrix}}{s+2}, \\
P_2(s)&= \frac{\begin{bmatrix}17&7\\5&26\end{bmatrix}}{s} \!+\! \frac{\begin{bmatrix}14s\!-8 & 6s\!-10\\12s\!-14 & 6s\!-8\end{bmatrix}}{s^2+1} \!+\! \frac{\begin{bmatrix} 1& s+4 \\ s+5& (s\!+\!3)(s\!+\!1) \end{bmatrix}}{(s+6)(s+2)}, \\
P_3(s)&= \frac{\begin{bmatrix}14&17\\26&34\end{bmatrix}}{s} \!+\! \frac{\begin{bmatrix}8s-4 & 8s-8\\8s-2 & 2s-4\end{bmatrix}}{s^2+1} \!+\! \frac{10\!\!\begin{bmatrix} (s\!+\!8)(s\!+\!3)& s\!+\!14 \\ s-5& s\!+\!7 \end{bmatrix}}{(s+4)(s+2)}, \\
P_4(s)&= \frac{\begin{bmatrix}4&3\\2&13\end{bmatrix}}{s} + \frac{\begin{bmatrix}6s & 6s-8\\6s-22 & 2s-8\end{bmatrix}}{s^2+1} \\
&\qquad \ \ + \frac{4\begin{bmatrix} s+10& (s+4)(s+8) \\ (s+6)(s+2)(s+20)& s+3 \end{bmatrix}}{(s+11)(s+20)(s+5)}, \\
P_5(s)&= \frac{\begin{bmatrix}2&2\\7&13\end{bmatrix}}{s} + \frac{\begin{bmatrix}2s-4 & -12\\2s-8 & 2s-10\end{bmatrix}}{s^2+1} + \frac{\begin{bmatrix} 20(s+6)& s+4 \\ s-5& 3 \end{bmatrix}}{s+9}.
\end{split}
\end{equation*}
\normalsize
The agents have common imaginary-axis poles $j\Omega=\{0,\pm j1\}$.
The network topology is illustrated in Figure \ref{fig:syn_design}, which has two strongly connected components with $\phi_{\ess}(L_{11})=0.5236$ and $\phi_{\ess}(L_{22})=0$. By solving LMIs \eqref{eq:LMI_uniform}, we obtain $K_0, K_1$ so that conditions in Theorem~\ref{thm:synbility} are satisfied. Then, a uniform controller is designed by using the approach proposed in Section \ref{synuniform}, which is given by (25). The agents finally reach synchronization, as shown in Figure \ref{fig:syn_uniform}.

\newcounter{TempEqCnt}
\setcounter{TempEqCnt}{\value{equation}}
\setcounter{equation}{24}
\begin{figure*}[ht]
\hrulefill
\scriptsize
\begin{equation}
C(s)\!=\!0.01\frac{\begin{bmatrix}
62.14s^6 \!+\! 251.9s^5 \!+\! 444s^4 \!+\! 499.6s^3 \!+\! 422s^2 \!+\! 234.4s \!+\! 57.76  &  -42.5s^6 \!-\! 157s^5 \!-\! 240.3s^4 \!-\! 241.6s^3 \!-\! 215.1 s^2 \!-\! 136.8 s \!-\! 37.47 \\
\!-71.2 s^6 \!-\! 276.3 s^5 \!-\! 463.7 s^4 \!-\! 516 s^3 \!-\! 460.6 s^2 \!-\! 273.8 s \!-\! 70.57 & 67.56 s^6 \!+ 277.3 s^5 \!+ 500 s^4 \!+ 578 s^3 \!+ 493.7 s^2 \!+ 272.2 s \!+ 66.28
\end{bmatrix}}{16 s^6 + 96 s^5 + 240 s^4 + 320 s^3 + 240 s^2 + 96 s + 16}
\end{equation}
\end{figure*}
\normalsize

\section{Conclusions}
\label{conclusion}


In this paper, we studied the synchronization of dynamical networks from a novel phasic perspective. For the analysis problem, we obtained scalable synchronization conditions in the form of small phase type inequalities. For the synthesis part, we characterized the largest diversity of the agents that can be tolerated without undermining the network synchronizability in terms of phase bounded cones. We provided design algorithms of synchronizing controllers under the cases of agent-dependent and uniform controllers. We wish to convey through these new results a message that phase gives a unique opportunity in unraveling hidden facts on synchronization that have not been revealed via traditional approaches.

For synthesis problem, while requiring a uniform controller is favorable in terms of scalability, it may happen that this poses too much restriction so that the problem becomes unsolvable. Shifting to an agent-based controller design will rectify the problem, but at the cost of losing all the scalability. We are interested in finding a solution that can strike a balance between conservatism and scalability. Using cluster-based controller design is an option.

Note that the solvability conditions in the case of uniform controller essentially require the simultaneous sectorization of the residue matrices of different agents at the imaginary-axis poles. If the agents can be divided into several clusters so that the residue matrices of the agents within the same cluster have similar phases, then it is reasonable to design a common controller for the agents from the same cluster and allow different controllers to be designed for different clusters. This way, one can achieve a design with low conservatism and meanwhile maintain the scalability as much as possible.
Then, it is meaningful to investigate the agent clustering problem.




\begin{thebibliography}{999}

\bibitem{RenBeard08}
W.~Ren and R.~W.~Beard, {\em Distributed Consensus in Multi-vehicle Cooperative Control}, Springer, London, 2008.

\bibitem{Lin08}
Z.\ Lin, \emph{Distributed Control and Analysis of Coupled Cell Systems}, VDM Verlag Dr. Muller, Germany, May 2008.

\bibitem{BulloCortesMartinez} F.~Bullo, J. Cort\'{e}s, and S. Mart\'{i}nez, {\em Distributed Control of Robotic Networks}, Princeton University Press, 2009.

\bibitem{Mesbahi} M. Mesbahi and M. Egerstedt, {\em Graph Theoretic Methods in Multiagent Networks}, Princeton University Press, 2010.

\bibitem{Lewis14}
F.~L.~Lewis, H.~Zhang, K.~Hengster-Movric, and A.~Das, \emph{Cooperative Control of Multi-Agent Systems: Optimal and Adaptive Design Approaches}, Springer-Verlag, 2014.

\bibitem{Bullo} F.~Bullo, {\em Lectures on Network Systems}, ed. 1.4, Kindle Direct Publishing, 2019, with contributions by J. Cort\'{e}s, F. D\"{o}rfler, and S. Mart\'{i}nez.

\bibitem{SaberMurray04}
R.~Olfati-Saber and R.~M.~Murray, ``Consensus problems in networks of agents with switching topology and time-delays,'' {\em IEEE Trans. Autom. Control}, vol.~49, no.~9, pp.~1520-1533, 2004.

\bibitem{RenBeard05TAC}
W.~Ren and R.~W.~Beard, ``Consensus seeking in multiagent systems under dynamically changing interaction topologies,'' {\em IEEE Trans. Autom. Control}, vol.~50, no.~5, pp.~655-661, 2005.

\bibitem{Allgower2014}
M.~B{\"u}rger, D.~Zelazo, and F.~Allg{\"o}wer, ``Duality and network theory in passivity-based cooperative control,'' {\em Automatica}, vol.~50, no.~8, pp.~2051-2061, 2014.

\bibitem{HaraPassivity11}
N.~Fujimori, L.~Liu, and S.~Hara, ``Passivity-based hierarchical consensus for nonlinear multi-agent systems,'' in {\em Proc. SICE Annual Conference 2011}, pp.~750-753, 2011.

\bibitem{NIConsensus}
J.~Wang, A.~Lanzon, and I.~R.~Petersen, ``Robust output feedback consensus for networked negative-imaginary systems,'' {\em IEEE Trans. Autom. Control}, vol.~60, no.~9, pp.~2547-2552, 2015.

\bibitem{LestasVinnicombe10} I.~Lestas and G.~Vinnicombe, ``Heterogeneity and scalability in group agreement protocols: Beyond small gain and passivity approaches,'' {\em Automatica}, vol.~46, no.~7, pp.~1141-1151, 2010.

\bibitem{Khong16}
S.~Z.~Khong, E.~Lovisari, and A.~Rantzer, ``A unifying framework for robust synchronization of heterogeneous networks via integral quadratic constraints,'' {\em IEEE Trans. Autom. Control}, vol.~61, no.~5, pp.~1297-1309, 2016.

\bibitem{Morse2003} A.~Jadbabaie, J.~Lin, and A.~S.~Morse, ``Coordination of groups of mobile autonomous agents using nearest neighbor rules,'' {\em IEEE Trans. Autom. Control}, vol.~48, no.~6, pp.~988-1001, 2003.

\bibitem{RenBeard05ACC}
W.~Ren, R.~W.~Beard, and E.~M.~Atkins, ``A survey of consensus problems in multi-agent coordination,'' in {\em Proc. Amer. Control Conf.}, pp.~1859-1864, 2005.

\bibitem{LinFrancisMaggiore06}
Z.~Lin, B.~Francis, and M.~Maggiore, ``Getting mobile autonomous robots to rendezvous,'' in {\em Workshop on Control of Uncertain Systems:  Modelling, Approximation, and Design}, pp.~119-137, 2006.

\bibitem{SaberMurray} R.~Olfati-Saber, J.~A.~Fax, and R.~M.~Murray, ``Consensus and cooperation in networked multi-agent systems,'' {\em Proceedings of the IEEE}, vol.~95, no.~1, pp.~215-233, 2007.

\bibitem{FaxMurray} J.~A.~Fax and R.~M.~Murray, ``Information flow and cooperative control
of vehicle formations,'' {\em IEEE Trans. Autom. Control}, vol.~49, no.~9,
  pp.~1465-1476, 2004.

\bibitem{Ma}
C.-Q.~Ma and J.-F.~Zhang, ``Necessary and sufficient conditions for consensusability of linear multi-agent systems,'' {\em IEEE Trans. Autom. Control}, vol.~55, no.~5, pp.~1263-1268, 2010.

\bibitem{LiDuanChenHuang}
Z.~Li, Z.~Duan, G.~Chen, and L.~Huang, ``Consensus of multiagent systems and synchronization of complex networks: A unified viewpoint,'' {\em IEEE Trans. Circuits Syst. {I}, {R}eg. {P}apers}, vol.~57, no.~1, pp.~213-224, 2010.

\bibitem{YouXie}
K.~You and L.~Xie, ``Network topology and communication data rate for consensusability of discrete-time multi-agent systems,'' {\em IEEE Trans. Autom. Control}, vol.~56, no.~10, pp.~2262-2275, 2011.

\bibitem{GuLewis} G.\ Gu, L.~Marinovici, and F.~ L.~Lewis, ``Consensusability of discrete-time
dynamic multiagent systems,'' {\em IEEE Trans. Autom. Control}, vol.~57, no.~8, pp.~2085-2089, 2012.

\bibitem{WielandAllgower}
P.~Wieland, R.~Sepulchre, and F.~Allg{\"o}wer, ``An internal model principle is necessary and sufficient for linear output synchronization,'' {\em Automatica}, vol.~47, no.~5, pp.~1068-1074, 2011.

\bibitem{LuLiuFeng} M.~Lu, L.~Liu and G.\ Feng, ``Output synchronization of heterogeneous linear multi-agent systems,'' in {\em Proc. 11th Asian Control Conference,} pp.\ 156-161, 2017.

\bibitem{WCQ20}
D.~Wang, W.~Chen, and L.~Qiu, ``Synchronization of heterogeneous dynamical networks via phase analysis,'' \emph{21st IFAC World Congress}, pp.~3075-3080, 2020.

\bibitem{Hara_glocal}
S.\ Hara, H.\ Tanaka, and T.\ Iwasaki,
``Stability analysis of systems with generalized
frequency variables,'' {\em IEEE Trans. Autom. Control}, vol.~59, no.~2, pp.~313-326, 2014.

\bibitem{gu_qiu}
L.~D.~Alvergue, G.~Gu, and L.~Qiu, ``Output consensus control for multi-agent systems in the presence of gap metric uncertainties,'' in {\em Proc. 54th IEEE Conf. Decis. Control}, pp.~5581-5586, 2015.

\bibitem{CWKQcdc19}
W.~Chen, D.~Wang, S.~Z.~Khong, and L.~Qiu, ``Phase analysis of MIMO LTI systems,'' in \emph{Proc. 58th IEEE Conf. Decis. Control}, pp.~6062-6067, 2019.

\bibitem{CWKQ2021}
W.~Chen, D.~Wang, S.~Z.~Khong, and L.~Qiu, ``A phase theory of MIMO LTI systems,'' {\em arXiv preprint arXiv:2105.03630v2}, 2021.

\bibitem{laplacian-survey}
R.~Merris, ``Laplacian matrices of graphs: {A} survey,'' {\em Linear Algebra Appl.}, vol.~197, pp.~143-176, 1994.

\bibitem{WCKQ20}
D.\ Wang, W.\ Chen, S.\ Z.\ Khong, and L.\ Qiu, ``On the phases of a complex matrix,'' \emph{Linear Algebra Appl.}, vol.~593, pp.~152-179, 2020.

\bibitem{QWMC22}
L. Qiu, D. Wang, X. Mao, and W. Chen, ``On the phases of a semi-sectorial matrix,'' \emph{arXiv preprint arXiv:2205.07607}, 2022.

\bibitem{horntopics}
R.~A.~Horn and C.~R.~Johnson, {\em Topics in Matrix Analysis}, Cambridge
  University Press, 1991.

\bibitem{ZhangFuzhen2015}
F.~Zhang, ``A matrix decomposition and its applications,'' {\em Linear Multilinear Algebra}, vol.~63, pp.~2033-2042, 2015.

\bibitem{FurtadoJohnson2001}
S.~Furtado and C.~R.~Johnson, ``Spectral variation under congruence,'' {\em Linear Multilinear Algebra}, vol.~49, pp.~243-259, 2001.

\bibitem{FurtadoJohnson2003}
S.~Furtado and C.~R.~Johnson, ``Spectral variation under congruence for a
  nonsingular matrix with 0 on the boundary of its field of values,'' {\em Linear
  Algebra Appl.}, vol.~359, pp.~67-78, 2003.





\bibitem{Safonov1982}
M. Safonov, ``Stability margins of diagonally perturbed multivariable feedback
systems,'' {\em IEE Proceedings}, vol. 129, pt D, no. 6, pp. 251-256, 1982.

\bibitem{Bauer}
F. L. Bauer, ``Optimal scaled matrices,'' {\em Numer. Math.}, vol. 5, pp. 73-87, 1963.

\bibitem{Doyle1982}
J. C. Doyle, ``Analysis of feedback systems with structured uncertainties,'' {\em IEE
Proceedings}, vol. 129, pt D, no. 6, pp. 242-250, 1982.

\bibitem{Zhou}
K.~Zhou, J.~C.~Doyle, and K.~Glover, {\em Robust and Optimal Control},
  Prentice Hall, New Jersey, 1996.

\bibitem{StoerWitzgall}
J.\ Stoer and C.\ Witzgall, ``Transformations by diagonal matrices in a normed space,'' {\em Numer. Math.}, vol.~4, pp. 158-171, 1962.


\bibitem{BrualdiRyser}
R.~A.~Brualdi and H.~J.~Ryser, {\em Combinatorial Matrix Theory}, Cambridge University Press, 1991.

\bibitem{Kottenstette}
N.~Kottenstette, M.~J.~McCourt, M.~Xia, V.~Gupta, and P.~J.~Antsaklis, ``On
  relationships among passivity, positive realness, and dissipativity in linear
  systems,'' \emph{Automatica}, vol.~50, no.~4, pp.~1003-1016, 2014.

\bibitem{LiuYao16}
K.-Z.~Liu and Y.~Yao, {\em Robust Control: Theory and Applications}, John Wiley \& Sons, 2016.


\bibitem{SuHuang}
Y.~Su and J.~Huang, ``Cooperative output regulation of linear multi-agent systems by output feedback,'' {\em Syst. Control Lett.}, vol.~61, pp.~1248-1253, 2012.




\bibitem{Lagrange-interpolation}
H.~Jeffreys and B.~S.~Jeffreys, ``Lagrange's interpolation formula,'' \S9.011, in {\em Methods of Mathematical Physics}, 3rd ed., Cambridge University Press, Cambridge, p.~260, 1988.



\end{thebibliography}

\appendix

\section{A Major Technical Lemma}

Let the agents be divided into groups of the form \eqref{eq: agents_decomp} and the agent-dependent controllers be grouped accordingly in the form of (\ref{eq: controller_decomp}). The Laplacian matrix $L_0$ is written in the form of \eqref{eq: L_decomp}. Let $C_i(s)=\epsilon \tilde{C}_i(s)$, where $\epsilon>0$ is a gain to be chosen as needed. Denote
\begin{align*}
\tilde{C}_i(0)=K_{0i}\in\mathbb{R}^{m\times m} \text{ and }\tilde{C}_i(j\omega_k)=K_{ki}\in\mathbb{C}^{m\times m}
\end{align*}
for $i=1,\dots,n,k=1,\dots,q$. Let $\boldsymbol{M}_{k\kappa}$ be as in \eqref{Mkk} and denote
\begin{align*}
    \boldsymbol{K}_{k\kappa}&=\diag\{K_{k(n_{\kappa-1}+1)},\dots, K_{kn_{\kappa}}\},
\end{align*}
where $\kappa=1,\dots,\nu,k=0,\dots,q$ and $n_0=0$.

\begin{lemma}\label{mtl}
If matrices
$\!-\boldsymbol{M}_{k\kappa}\boldsymbol{K}_{k\kappa}(L_{\kappa\kappa}\!\otimes I_m)$ are Hurwitz for $\kappa=1,\dots,\nu,k=0,\dots,q$, then synchronization is reached with $C_i(s)\!=\!\epsilon\tilde{C}_i(s),i\!=\!1,\dots,n$ for all $\epsilon\!\in\!(0,\epsilon^*)$, where $\epsilon^*$ can be estimated from given data.
\end{lemma}
\begin{pf}
It suffices to show the Gang of Four matrices (\ref{gof1}) and (\ref{gof2}) are stable. We first show the stability of (\ref{gof1}).
Without loss of generality, assume $q=1$. The case when $q>1$ can be shown by the same approach with a dimension expansion.

Let $\hat{\boldsymbol{P}}(s)$ and $\hat{L}$ be as in (\ref{Phat}) and (\ref{Lhat}).
Then, simple computation yields
\begin{align*}
\hat{\boldsymbol{P}}(s)=&\epsilon(Q_1'\!\otimes\! I_m) \!\left( \frac{\boldsymbol{M}_{01}\boldsymbol{K}_{01}}{s} \!+\! \frac{\boldsymbol{M}_{11}\boldsymbol{K}_{11}}{s-j\omega_1}   \!+\! \frac{\bar{\boldsymbol{M}}_{11}\bar{\boldsymbol{K}}_{11}}{s+j\omega_1} \right) \\
&\cdot(V_1^{-1}Q_1\otimes I_m)
+\epsilon\tilde{\Delta}(s),
\end{align*}
where $\tilde{\Delta}(s)$ is stable.
Let $\sys{A_\Delta}{B_\Delta}{C_\Delta}{0}$ be a minimal realization of $\tilde{\Delta}(s)$. Then a realization of $\hat{\boldsymbol{P}}(s)$ is given by
\begin{align*}
\left[ \begin{array}{c|c}
\begin{bmatrix}0_{(n_1\!-1)m}\!& & & \\ \!&j\omega_1I_{(n_1\!-1)m}\! & & \\ &\! &\!-\!j\omega_1I_{(n_1\!-1)m}\!& \\& & &\!A_{\Delta}\end{bmatrix}\! & \epsilon\!\! \begin{bmatrix}  B_0 \\ B_1 \\ \bar{B}_1 \\ B_\Delta \end{bmatrix}\!\! \\ [10mm]\hline \\[-5mm] \begin{bmatrix}I_{(n_1-1)m} &\ \ \ \ I_{(n_1-1)m} \ &\ \ \ \ I_{(n_1-1)m}\ \ \ \ & C_\Delta\end{bmatrix}& 0
\end{array}\! \right]\!,
\end{align*}
where
\begin{align*}
B_0&=(Q_1'\!\otimes\! I_m)\boldsymbol{M}_{01}\boldsymbol{K}_{01}(V_1^{-1}Q_1\!\otimes\! I_m),\\
B_1&=(Q_1'\!\otimes\! I_m)\boldsymbol{M}_{11}\boldsymbol{K}_{11}(V_1^{-1}Q_1\!\otimes\! I_m).
\end{align*}
Thus, the state matrix of $\hat{\boldsymbol{P}}(s)\#\hat{L}$ is given by
\begin{align*}
N\!\!=\!\!\begin{bmatrix}0& & & \\ &j\omega_1I & & \\ & &-j\omega_1 I& \\ & & &A_{\Delta}\end{bmatrix}\!-\! \epsilon \!\begin{bmatrix}  B_0 \\B_1  \\\bar{B}_{1}  \\B_{\Delta} \end{bmatrix}\!(\hat{L}\!\otimes\! I_m)\! \begin{bmatrix}I & I& I & C_{\Delta}\!\end{bmatrix}\!.
\end{align*}
We will next show that $N$ is Hurwitz stable. According to the conditions stated in this lemma, we have
\begin{align*}
\angle \lambda \!\left(\! B_0(\hat{L}\!\otimes\! I_m) \!\right)\! \in\! (\!-\frac{\pi}{2}, \frac{\pi}{2}),\
\angle \lambda \! \left(\! B_{1}(\hat{L}\!\otimes\! I_m)\! \right) \!\in\! (\!-\frac{\pi}{2}, \frac{\pi}{2}).
\end{align*}
Therefore, there exist $X_0, X_1>0$ such that
\begin{align*}
X_0 B_0(\hat{L}\otimes I_m)+(\hat{L}' \otimes I_m)B_0^*X_0 &=I,\\
X_{1}B_{1}(\hat{L}\otimes I_m)+(\hat{L}' \otimes I_m)B_{1}^*X_{1}&=I,\\
\bar{X}_{1}\bar{B}_{1}(\hat{L}\otimes I_m) +(\hat{L}'\otimes I_m)\bar{B}_{1}^*\bar{X}_{1}&=I.
\end{align*}
Also, since $\tilde{\Delta}(s)$ is stable, there exists $X_{\Delta}>0$ such that $X_{\Delta}A_\Delta +A_\Delta^*X_{\Delta}=-I$. Now let
\begin{align*}
Y=\begin{bmatrix}
X_0 & \epsilon Y_{12} & \epsilon Y_{13} & 0\\
\epsilon Y_{12}^* & X_1 & \epsilon Y_{23} & 0\\
\epsilon Y_{13}^* & \epsilon Y_{23}^* & \bar{X}_1 & 0\\
0 & 0 & 0 & X_{\Delta}
\end{bmatrix},
\end{align*}
where
\begin{align*}
Y_{12}&=-\frac{j}{\omega_1}(X_0B_0(\hat{L}\otimes I_m)+(\hat{L}'\otimes I_m)B_1^*X_1),\\
Y_{13}&=\frac{j}{\omega_1}(X_0B_0(\hat{L}\otimes I_m)+(\hat{L}'\otimes I_m)\bar{B}_1^*\bar{X}_1),\\
Y_{23}&=\frac{j}{2\omega_1} (X_1B_1(\hat{L}\otimes I_m)+(\hat{L}\otimes I_m)\bar{B}_1^*\bar{X}_1).
\end{align*}
Some algebraic computation yields
\begin{align*}
YN+N^*Y=-\epsilon \left( \begin{bmatrix} I & S \\S^* & \frac{1}{\epsilon}I+T\end{bmatrix}+\epsilon Z \right),
\end{align*}
where $S,T$ and $Z$ do not depend on $\epsilon$.
One can see that there exists $\epsilon^*>0$ such that
\begin{align*}
    Y>0 \text{ and }\begin{bmatrix} I & S \\S^* & \frac{1}{\epsilon}I+T\end{bmatrix}+\epsilon Z>0
\end{align*}
for all $0<\epsilon<\epsilon^*$, meaning that $Y$ is a positive definite solution to the Lyapunov inequality
$YN+N^*Y<0$.
Therefore, $N$ is stable for all $0<\epsilon<\epsilon^*$. This shows the stability of (\ref{gof1}). The stability of (\ref{gof2}) can be shown similarly. \hfill\qed
\end{pf}

\end{document}